\documentclass{amsart}

\usepackage{amsmath}
\usepackage{amssymb}
\usepackage{amsfonts}
\usepackage{mathrsfs}
\usepackage{xspace}
\usepackage{bm}
\usepackage{fancyref}
\usepackage{textcomp}
\usepackage[Symbol]{upgreek}

\newcommand{\KLdiv}[2]{D ( #1 || #2 )}
\newcommand{\setD}{\mathcal{D}}
\newcommand{\setE}{\mathcal{E}}
\newcommand{\setL}{\mathcal{L}}
\newcommand{\setQ}{\mathcal{Q}}
\newcommand{\setS}{\mathcal{S}}
\newcommand{\setT}{\mathcal{T}}
\newcommand{\set}[1]{\left\{#1\right\}}
\newcommand{\Proba}[1]{\Prob \! \left[ #1 \right]}
\newcommand{\card}[1]{\lvert#1\rvert}
\newcommand{\naturals}{\mathbb N}
\newcommand{\elementof}[2]{\left[#1\right]_{#2}}
\newcommand{\Exop}{\mathbb{E}}
\newcommand{\Ex}[1]{\Exop \! \left[#1\right]}

\def\be{\begin{equation}}
\def\ee{\end{equation}}
\def\een{\nonumber \end{equation}}
\def\ba#1\ea{\begin{align}#1\end{align}}
\def\bas#1\eas{\begin{align*}#1\end{align*}}

\usepackage{graphicx}

\newcommand{\eps}{\varepsilon}
\newcommand{\RR}{\mathbb{R}}

\newcommand{\EE}{\mathbb{E}}

\newtheorem{theorem}{Theorem}[section]

\newtheorem{lemma}[theorem]{Lemma}

\newtheorem{remark}[theorem]{Remark}

\newcommand{\Prob}{\operatorname{Prob}}

\newcommand{\tr}{\operatorname{Tr}}
\newcommand{\rank}{\operatorname{Rank}}
\newcommand{\argmin}{\operatorname{argmin}}

\begin{document}

\title[Decoding binary node labels from censored edge measurements]{Decoding binary node labels from censored edge measurements: Phase transition and efficient recovery}

\author{Emmanuel~Abbe, 
        Afonso~S.~Bandeira, Annina~Bracher, and~Amit~Singer}% <-this % stops a space
\address{Emmanuel Abbe is with the Program in Applied and Computational Mathematics (PACM) and the Department of Electrical Engineering, Princeton University, Princeton, NJ 08544, USA. E-mail: {\tt eabbe@princeton.edu}.}% <-this % stops a space
\address{Afonso~S.~Bandeira is with PACM, Princeton University, Princeton, NJ 08544, USA. E-mail: {\tt ajsb@math.princeton.edu}.}% <-this % stops a space
\address{Annina Bracher is with the Department of Electrical Engineering, Swiss Federal Institute of Technology, Zurich, ZH 8092, CH. E-mail: {\tt bracher@isi.ee.ethz.ch}.}% <-this % stops a space
\address{Amit Singer is with the Department of Mathematics and PACM, Princeton University, Princeton, New Jersey 08544, USA. E-mail: {\tt amits@math.princeton.edu}.}% <-this % stops a space
\maketitle

\begin{abstract}
We consider the problem of clustering a graph $G$ into two communities by observing a subset of the vertex correlations.  
Specifically, we consider the inverse problem with observed variables $Y=B_G x \oplus Z$, where $B_G$ is the incidence matrix of a graph $G$,  
$x$ is the vector of unknown vertex variables (with a uniform prior), and $Z$ is a noise vector with Bernoulli$(\eps)$ i.i.d.\ entries. All variables and operations are Boolean. 
This model is motivated by coding, synchronization, and community detection problems. In particular, it corresponds to a stochastic block model or a correlation clustering problem with two communities and censored edges.  
Without noise, exact recovery (up to global flip) of $x$ is possible if and only the graph $G$ is connected, with a sharp threshold at the edge probability $\log(n)/n$ for Erd\H{o}s-R\'enyi random graphs. 
The first goal of this paper is to determine how the edge probability $p$ needs to scale to allow exact recovery in the presence of noise. Defining the degree rate of the graph by $\alpha =np/\log(n)$, it is shown that exact recovery is possible if and only if $\alpha >2/(1-2\eps)^2+ o(1/(1-2\eps)^2)$. In other words, $2/(1-2\eps)^2$ is the information theoretic threshold for exact recovery at low-SNR.  
In addition, an efficient recovery algorithm based on semidefinite programming is proposed and shown to succeed in the threshold regime up to twice the optimal rate. For a deterministic graph $G$, defining the degree rate as $\alpha=d/\log(n)$, where $d$ is the minimum degree of the graph, it is shown that the proposed method achieves the rate $\alpha> 4((1+\lambda)/(1-\lambda)^2)/(1-2\eps)^2+ o(1/(1-2\eps)^2)$, where $1-\lambda$ is the spectral gap of the graph $G$.
\\

A preliminary version of this paper appeared in ISIT 2014~\cite{conferenceversion}. This version will appear in the IEEE Transactions on Network Science and Engineering. 
\end{abstract}

%\begin{IEEEkeywords}
%Synchronization problem, Information theoretic bounds, Erd\H{o}s-R\'enyi graphs, Semidefinite relaxations, graph-based codes. 
%\end{IEEEkeywords}

% \fi
%\IEEEpeerreviewmaketitle

%\section{Introduction}
%% and "HIS" in caps to complete the first word.
%\IEEEPARstart{T}{his} demo file is intended to serve as a ``starter file''
%for IEEE journal papers produced under \LaTeX\ using
%IEEEtran.cls version 1.8 and later.
%% You must have at least 2 lines in the paragraph with the drop letter
%% (should never be an issue)
%I wish you the best of success.

\section{Introduction}
A large variety of problems in information theory, machine learning, and image processing are concerned with inverse problems on graphs, i.e., problems where a graphical structure governs the dependencies between the variables that are observed and the variables that are unknown. In simple cases, the dependency model is captured by an undirected graph with the unknown variables attached at the vertices and the observed variables attached at the edges. 
Let $G=(V,E)$ be a graph with vertex set $V$ and edge set $E$, and let $x^V$ be the vertex- and $y^E$ the edge-variables. In many cases of interest (detailed below), the probabilistic model for the edge-variables {\it conditionally} on the vertex-variables has a simple structure: it factorizes as    
\begin{align}
P(y^E|x^V)=\prod_{e \in E} Q(y_e|x[e]), \label{gene}
\end{align}
where $y_e$ denotes the variable attached to edge $e$, $x[e]$ denotes the two vertex-variables incident to edge $e$, and $Q$ is a local probability kernel. In this paper, we consider Boolean edge- and vertex-variables, and assume that the kernel $Q$ is symmetric and depends only on the XOR of the vertex-variables.\footnote{Symmetry means that $Q(y|x_1,x_2)=P(y|x_1 \oplus x_2)$ for some $P$ that satisfies $P(1|1)=P(0|0)$.} The edge-variables can then be viewed as a random vector $Y^E$ that satisfies
\begin{align}
Y^E=B_G x^V \oplus Z^E, \label{lin}
\end{align}
where $B_G$ is the incidence matrix of the graph, i.e., the $m\times n$ matrix, with $m=|E|$ and $n=|V|$, such that $B_G(e,v)=1$ if and only if edge $e$ is incident to vertex $v$, and $Z$ is a random vector of dimension $|E|$ representing the noise.

In the above setting, the forward problem of recovering the most likely edge-variables given the vertex-variables is trivial and amounts to maximizing $Q$ for each edge. The inverse problem, however, is more challenging: the most likely vertex-variables (say with a uniform prior) given the edge-variables cannot be found by local maximization.

This problem can be interpreted as a community detection problem with censored edges: Consider a population with $n$ vertices and two communities, the blues and the reds. The colors of the vertices, encoded by the binary variables $\{X_i\}_{i \in [n]}$, are unknown and the goal is to recover them by observing pairwise interactions of these nodes. However, not all ${n \choose 2}$ interactions are observed, only the ones encoded by the graph $G$.   
%if vertex $i$ and $j$ are connected by an edge of $G$, we get to observe the interaction between $X_i$ and $X_j$, otherwise no information is revealed. 
In the noiseless case, the observation is perfect and allows to determine whether $X_i$ and $X_j$ are in the same community or not, i.e., $Y_{ij}=X_i \oplus X_j$. Hence, recovering the partition in this case amounts to having a connected graph $G$, and the recovery is obtained by picking a vertex label and recovering the other vertices along any spanning tree. Note that we can only hope to recover the partition and not the exact colors, as a global flipping of all the colors gives the same observations. 
In the more interesting setting, the observations are assumed to be noisy, i.e., with probability $\eps$ an error is made on the parity of the two colors: $Y_{ij}=X_i \oplus X_j \oplus Z_{ij}$, where the $Z_{ij}$'s are i.i.d.\ Bernoulli$(\eps)$. In this case, the connectivity of $G$ is a necessary condition, but it is in general not sufficient to cope with the noise. This paper investigates how to strengthen the connectivity assumption, in terms of the edge probability for random graphs or in terms of the spectral gap for deterministic graphs, in order to recover the partition despite the noise.  

There are various interpretations and models that connect to this problem. 
\begin{itemize}
\item {\bf Community detection:} It is worth connecting the above model to other existing models for community networks. 
%{\bf Community detection:} 
%Various information theory\footnote{In a different context, information theoretic tools such as Fano's inequality were used in \cite{fano1} and \cite{fano2} to analyze model selection of graphical models.} concepts have been used to perform clusterings of deterministic graphs. 
%In \cite{rosvall1,sun,rosvall2}, a communication process is proposed to model the clustering of a graph, with the original graph as the input to a channel which produces as the output the synthesis of the clustering. The conditional entropy is then used to model the accuracy of the clustering, to be traded with its complexity, e.g., relying on MDL principles \cite{mdl1,mdl2}. 
%Similar approaches were proposed in \cite{ziv1}, based on the information bottleneck method \cite{bottle} and relying on mutual information for accuracy. We refer to \cite{fortunato,newman-girvan,airoldi,newman} for surveys on these models.  
%In all the above works, the mutual information or conditional entropy is used to estimate the information among graphs, in particular between the original and clustered graphs, or to perform model selection. 
%, with a large scope of applications in image processing \cite{image1,image2}, social networks \cite{social1} and biological networks \cite{newman-girvan,fortunato}. 
%While hypergraphs are relevant for coding theory applications, graphs are natural to model networks. 
The model in \eqref{gene} can be seen as a general probabilistic model of networks, that extends the   
basic Erd\H{o}s-R\'enyi model \cite{ER-seminal}, %where edge-variables are drawn from an i.i.d. ensemble, 
which often turns out to be too simplistic since all vertices have the same expected degree and no cluster structure appears. One possibility to obtain cluster structure is precisely to attach latent variables to the vertices and assume an edge distribution that depends on these variables. There are various models with latent variables, such as the exchangeable, inhomogeneous or stochastic block models \cite{airoldi,sbm1,sbm2,dyer,newman2,sbm-book}.
The general model in \eqref{gene} can be used for this purpose, as explained above in the special case of \eqref{lin}.   
The vertex-variables represent the community assignment, the edge-variables the connectivity, and the graph $G$ encodes where the information is available. 
%In the noiseless additive case and for every edge $e=(i,j)$, the edge-variable $y_{e}=x_i \oplus x_j$ encodes whether the vertices $i$ and $j$ are in the same community or not. 
%With noise on top, vertices in the same/different communities are also allowed disconnect/connect, which is a more realistic model. 
%A more interesting model is to allow vertices in the same/different communities to also disconnect/connect, corresponding to a noisy version like \eqref{gene}.% and 
% %Consider the case of a network with two communities, and label each vertex of the graph with binary variables encoding the communities, which are assumed for example to have roughly the same size. 
% %Consider now the following network model: There is a graph $G$ which describes for each pair of nodes whether the information about their connectivity is available. When the information is available, the pair of vertices are assumed to be connected with a probability that depends on their labels. 
% %Hence, for each pair of vertices there are three possible outcomes: (i) no information available, (ii) the vertices are connected, (iii) the vertices are not connected. 
% the community detection problem is the inverse problem discussed above. 
%\end{itemize}
The model \eqref{lin} is related to the stochastic block model through the following  {\bf censored block model}, introduced in \cite{random} in a different context. Given a base-graph $G=(V,E(G))$ and a community assignment $X \in \{0,1\}^V$, the following random graph is generated on the vertex set $V$ with ternary edge labels $E_{ij} \in \{*,0,1\}$ drawn independently with the following probability distribution: 
\begin{subequations}\label{bl:filBlMod}  
\begin{align}
&\mathbb{P}\{ E_{ij} = * | E(G)_{ij}=0 \} =1\\
%&\quad = \mathbb{P}\{ E_{ij} = * | X_i \neq X_j , E(G)_{ij}=0\}=1,\\
&\mathbb{P}\{ E_{ij} = 1 | X_i=X_j , E(G)_{ij}=1\} = q_1,\\
&\mathbb{P}\{ E_{ij} = 1 | X_i \neq X_j , E(G)_{ij}=1\} = q_2.
\end{align} 
\end{subequations}
%In particular, when $X$ is uniformly distributed over $\{0,1\}^V$, the communities are balanced. %and the graph ensemble is denoted $\mathcal{G}(n,q_1,q_2,G)$.  
Put differently, \eqref{bl:filBlMod} is a graph model where information is only available on the base-graph $G$, the $*$-variable encodes the absence of information, and when information is available, two vertices are connected with probability $q_1$ if they are in the same community and with probability $q_2$ if they are in different communities. When $G=K_n$ is the complete graph and $X$ is uniformly distributed, this is the standard stochastic block model with two communities, and $q_1=a/n$, $q_2=b/n$ gives the sparse regime of \cite{decelle,mossel-sbm}. 
%The logarithmic degree regime was recently investigate in~\cite{Abbe_ExactSBM}. 
In the case of \eqref{lin}, the linear structure implies $q_1=1-q_2=\varepsilon$, which may be both of order 1, whereas the base-graph may be sparse. This raises an important distinction: in the sparse stochastic block model, it is assumed that most node pairs are unlikely to be connected, whereas in the model of this paper, it is assumed that information is not available for most node pairs. These are not the same, and the latter may help preventing false-alarm type of errors. However, we restrict ourselves in this paper to the symmetric case $q_1=1-q_2=\varepsilon$, which simplifies the computations. 
%The non-symmetric case was recently studied, with similar techniques, in~\cite{Abbe_ExactSBM}.
%The inverse problems in \eqref{gene} and \eqref{lin} arise also in various other contexts: 
%\begin{itemize}

\item {\bf Correlation clustering:} \cite{corr_cluster} considers the problem of clustering a complete graph with edges labeled in $\{-,+\}$ in order to maximize the number of agreeing edges (having a $+$ label within a cluster and a $-$ label otherwise). Another variant is proposed in \cite{nicolo}. The original motivation behind correlation clustering is to let the number of clusters be a design parameter, although the case of constraining the number of clusters has also been considered \cite{corr_cluster_k}. In our setting, the number of clusters is fixed and assumed to be 2. More importantly, our goal is to understand how sparse %{\it sub-complete}
the measurement graph can be in order to still be able to recover the original clustering, which is planted.    
In that regard, we are proposing a planted correlation clustering problem with a fixed number of clusters, censored measurements, and with a probabilistic model.

\item {\bf Coding:} Equation \eqref{lin} provides the output on a binary symmetric channel of a code whose generator matrix is the adjacency matrix of the graph $G$. More precisely, since here $G$ is assumed to be a graph and not a hyper-graph, this is a very simple code, namely a 2-right-degree LDGM code. While this is not a particularly interesting code by itself (e.g., at any fixed rate, it has a constant fraction of isolated vertices), it is a relevant primitive for the construction of other codes such as LT or raptor codes \cite{kumar,LT-shokro}. Note that this paper will consider such a code at a vanishing rate, namely $c/\log(n)$, and determine for which values of $c$ the successful decoding of this code is still possible. Somehow unexpectedly, the Shannon capacity will also arise in this regime as shown in our main results.  

\item {\bf Constraint satisfaction problems:} \eqref{gene} is a particular case of the graphical channel studied in \cite{random} in the context of hypergraphs. 
This class of models allows in particular to recover instances of planted constraint satisfaction problems (CSPs) by choosing uniform kernels $Q$, where the vertex-variables represent the planted assignment and the edge-variables represent the clauses. In the case of a simple graph and not a hypergraph, this provides a model for planted formulae such as 2-XORSAT (model \eqref{lin}).

\item {\bf Synchronization:} 
Equation \eqref{lin} results also from the synchronization problem studied in \cite{ASinger_2011_angsync,Bandeira_Singer_Spielman_OdCheeger,Wang_RobustSynchronization,Alexeev_PhaseRetrievalPolarization,boumal}, if the dimension is one (e.g., when each vertex-variable is the 1-bit quantization of the reflection of a signal). The goal in synchronization over $O(r)$, the group of orthogonal matrices\footnote{Note that $O(r)$ denotes the group of orthogonal matrices of size $r\times r$ and does not refer to the big-O notation frequently used in algorithm analysis.} of size $r\times r$, is to recover the original values of the node-variables $\{x_j\}_{j \in [n]}$ in $O(r)$ given the relative measurements $\{Z_{ij} x_i^{-1} x_j\}_{i,j \in [n]}$, where $Z_{ij}$ is randomly drawn in $O(r)$ if the vertices $i$ and $j$ are adjacent and all-zero otherwise.\footnote{If $Z_{ij}$ is the $r \times r$ identity matrix, then the measurement is noise-free.} When $r=1$, we have $O(1)=\{-1,+1\}$ and the synchronization problem is equivalent to \eqref{lin}. 
\end{itemize}

While the above mentioned problems are all concerned with related inverse problems on graphs, there are various recovery goals that can be considered. This paper focuses on {\it exact recovery}, which requires all vertex-variables to be recovered simultaneously with high probability as the number of vertices diverges. The probability measure may depend on the graph ensemble or simply on the kernel $Q$ if the graph is deterministic. Note, as mentioned previously, that exact recovery of all variables in the model \eqref{lin} is not quite possible: the vertex-variables $x^V$ and $1^V \oplus x^V$ produce the same output $Y^E$. Exact recovery is meant ``up to a global flipping of the variables''. %This does not occur with hyperedges of odd order.  
For {\it partial recovery,} only a strictly dominant constant fraction of the vertex-variables are to be recovered correctly with high probability as the number of vertices diverges. Put differently, the true assignment need only be positively correlated with the reconstruction.\footnote{We have recently became aware that~\cite{Heimlicher_SBM} studies partial recovery for the model of this paper.} 
The recovery requirements vary with the applications, e.g., exact recovery is typically required in coding theory to ensure reliable  communication, %It is also required in the synchronization problems.
while both exact and partial recovery are of interest in community detection problems. 
% is the best one can hope for sparse networks, while exact recovery can be considered for slightly denser graphs with logarithmic degree \cite{mcsherry}.

This paper focuses on exact recovery for the linear model \eqref{lin} with  Boolean variables, and on random Erd\H{o}s-R\'enyi and deterministic base-graphs $G$. For this setup, we identify the information theoretic (IT) phase transition for exact recovery in terms of the edge density of the graph and the noise level and devise an efficient algorithm based on semidefinite programming (SDP), which approaches the threshold up to a factor of $2$ in the Erd\H{o}s-R\'enyi case.
This SDP based method was first proposed in \cite{ASinger_2011_angsync}, and it shares many aspects with the SDP methods in several other problems \cite{So_MIMO,Huang_Guibas_Graphics}. %Interestingly, the exact recovery conditions we obtain are related to the independently obtained ones of \cite{Huang_Guibas_Graphics} in the context of consistent shape map estimation in computer graphics (see Remark \ref{abouttheorem51inHuangGuibas}).
%Our specific model relates to the problems discussed above as follows: In terms of coding, equation \eqref{lin2} describes the output of a binary symmetric channel precoded with a code whose generator matrix is equal to the transposed incidence matrix of the graph. In the case of Erd\H{o}s-R\'enyi graphs, this will be equivalent to a constant 2-right-degree LDGM code. As noted in \cite{kumar,LT-shokro}, such a code cannot achieve reliable communication: If the transmission rate is positive, then a constant fraction of information bits are not transmitted, since there are almost surely isolated vertices in the graph. However, the code structure is interesting when combined with similar additional layers as in raptor codes \cite{LT-shokro}. In terms of CSP, equation \eqref{lin2} is a ``noisy'' 2-XORSAT formula \cite{random}. In terms of synchronization, it corresponds to the case $d=1$ with measurement graph $G$. Finally, 

\section{Related work}\label{abouttheorem51inHuangGuibas}

While writing this paper we became aware of various exciting related work that was being independently developed:

A similar exact recovery sufficient condition, as (\ref{conditionneededdual}) for the SDP, was independently obtained by Huang and Guibas~\cite{Huang_Guibas_Graphics} in the context of consistent shape map estimation (see Theorem 5.1. in \cite{Huang_Guibas_Graphics}). Their analysis goes on to show, essentially, that as long as the probability of a wrong edge is a constant strictly smaller than $\frac12$, the probability of exact recovery converges to $1$ as the size of the graph is arbitrarily large. In the context of our particular problem, that claim was also shown in~\cite{Wang_RobustSynchronization}. Later, this analysis was improved by Chen, Huang, and Guibas~\cite{Chen_Huang_Guibas_Graphics} and, when restricted to our setting, it includes guarantees on the rates at which this phase transition happens. However, these rates are, to the best of our knowledge, only optimal up to polylog factors. On the other hand, we are able to show near tight rates. For a given $\epsilon$ that is arbitrarily close to $\frac12$ we give an essentially-tight bound (off by at most a factor of $2$) on the size of the graph and edge density needed for exact recovery (Theorem~\ref{SDP_ErdosReinyi}). To the best of our knowledge, our Theorem~\ref{SDPtheoremfordregular} is the only available result for deterministic graphs.

On the IT side, both converse and direct guarantees were independently obtained by Chen and Goldsmith~\cite{Chen_Goldsmith_ISIT2014}. However, while considering a more general problem, the results they obtain are only optimal up to polylog factors.

\section{Model and results}
In this paper, we focus on the linear Boolean model
\begin{align}
Y^E=B_G x^V \oplus Z^E, \label{lin2}
\end{align}
where the vector components are in $\{0,1\}$ and the addition is modulo 2. We require exact recovery for $x^V$ and consider for the underlying graph $G=(V,E)$, with $V=[n]$, both the Erd\H{o}s-R\'enyi model $\mathrm{ER}(n,p)$  where the edges are drawn i.i.d.\ with probability $p$, and deterministic $d$-regular graphs. 
We assume that the noise vector $Z^E$ has i.i.d.\ components, equal to 1 with probability $\varepsilon$. We assume\footnote{The noise model is assumed to be known, hence the regime $\eps \in [1/2,1]$ can be handled by adding an all-one vector to $Y^E$.}
 w.l.o.g.\ that $\varepsilon\in [0,1/2]$, where $\eps = 0$ means no noise (and exact recovery amounts to having a connected graph) and $\eps = 1/2$ means maximal noise (and exact recovery is impossible no matter how connected the graph is).
 The prior on $x^V$ is assumed to be uniform. 
Note that the inverse problem would be much easier if the noise model caused erasures with probability $\varepsilon$, instead of errors. Exact recovery would then be possible if and only if the graph was still connected after the noisy edges had been erased. 
Since there is a sharp threshold for connectedness at $p=\frac{\log(n)}{n}$, this would happen a.a.s.\ if $p=\frac{(1+\delta)\log(n)}{n(1-\varepsilon)}$ for some $\delta>0$. 
Hence $1/(1-\eps)$ is a sharp threshold in $np/\log(n)$ for the exact recovery problem with erasures and base-graph $\mathrm{ER}(n,p)$.

The goal of this paper is to find the replacement to the erasure threshold $1-\eps$ for the setting where the noise causes errors. Similarly to channel coding where the Shannon capacity of the BSC$(\varepsilon)$ differs from the BEC$(\varepsilon)$ capacity, we obtain for the considered inverse problem the expression 
\begin{align}
D(1/2||\eps) & = (1-2\eps)^2/2 + o((1-2\eps)^2) \nonumber \\ & =\log(2)-H(\eps)+o((1-2\eps)^2),
\end{align}
where $D(1/2||\eps)$ is the Kullback-Leibler divergence\footnote{All logarithms have base $e$, i.e., we denote by $\KLdiv{1/2}{\varepsilon}=1/2 \log(1/(2\eps))+1/2 \log(1/(2(1-\eps)))$ the Kullback-Leibler divergence between $1/2$ and $\varepsilon$ and by $H \! \left( \varepsilon \right) = \varepsilon \log (1/\varepsilon) + (1-\varepsilon) \log (1/(1-\varepsilon))$ the entropy (in nats) of a binary random variable that assumes the value $1$ with probability $\varepsilon \in \left[0,1\right]$.} between $1/2$ and $\varepsilon$. Hence the Shannon capacity provides the threshold for the low-SNR regime, although the considered inverse problem is a priori not related to the channel coding theorem. 

More precisely, this paper establishes an IT necessary condition that holds for every graph (Theorem~\ref{th:convKLDiv}), an IT sufficient condition for Erd\H{o}s-R\'enyi graphs (Theorem~\ref{th:suffCondErdRen}), and an IT sufficient condition that holds for any graph (Theorem~\ref{IT_sufficient_CheegerConstant}) and depends on the graph's Cheeger constant, a common measure of the connectivity of a graph (see (\ref{def:Cheegerconstant44})) related to its spectral gap by Cheeger's inequality (see Theorem~\ref{thm:Cheegersineq}). Moreover, we also give a recovery guarantee that holds for an efficient algorithm based on SDP (Theorems~\ref{SDP_ErdosReinyi} and~\ref{SDPtheoremfordregular}).

In particular, we show that, for $\eps \to \frac12$ and $\frac12-\eps = \Omega \! \left( n^{-\tau} \right)$ for every $\tau > 0$: The bounds for the necessary condition for a general graph and the IT sufficient condition for the Erd\H{o}s-R\'enyi graph match.\footnote{The regime $\eps \to \frac12$ is frequently studied in the synchronization problem in dimension $d = 1$.} Remarkably, the sufficient condition for the efficient SDP-based method to achieve exact recovery matches the IT bound up to a factor of $2$.

If the noise parameter $\varepsilon$ is bounded away from both zero and $1/2$, then all conditions imply $d = \Theta \! \left( \log \! \left( n \right) \right)$, where $d$ is the expected average degree: $d = pn$. 
The factors by which the bounds differ decrease with an increasing noise parameter $\varepsilon$. Since in the noise-free case exact recovery is possible if and only if the graph is connected, which is true for trees (with $d \leq 2$) and, for Erd\H{o}s-R\'enyi graphs only when $d \geq \log \! \left( n \right)$, the factors between the necessary condition and the sufficient conditions necessarily approach infinity when $\varepsilon$ decreases to zero (since $D(1/2||\eps)$ diverges).

\section{Information Theoretic Bounds}

This section presents necessary and sufficient conditions for exact recovery of the vertex-variables $x^V$ from the edge-variables $Y^E$. We speak of exact recovery if there is a decoding algorithm that recovers the vertex-variables $x^V$ up to an unavoidable additive offset $\phi \in \left\{ 0^V, 1^V \right\}$ with some probability that converges to $1$ as the number of vertices approaches infinity.

By definition, maximum a posteriori (MAP) decoding always maximizes the probability of recovering the correct vertex-variables. Since we assume uniform priors, maximum likelihood (ML) and MAP decoding coincide. Hence, our definition of exact recovery is tantamount to requiring that ML decoding recovers the vertex-variables $x^V$ up to an unavoidable additive offset $\phi \in \left\{ 0^V, 1^V \right\}$ with some probability that converges to $1$ as the number of vertices approaches infinity. Note that an ML decoder produces vertex-variables $\tilde x^V$ that minimize the number of edges $(i,j)$ of $G$ for which  $Y^E_{ij} \oplus \tilde x_i \oplus \tilde x_j$ is non-zero.
%\[
%\min_{x^V_i\in\{0,1\}} \sum_{(i,j)\in G} \left| Y^E_{ij} \oplus x^V_i \oplus x^V_j  \right|.
%\]

\subsection{A Necessary Condition for Successful Recovery}

For each graph $G = \left( V, E \right)$ (drawn from the Erd\H{o}s-R\'enyi model or not), the following result holds:

\begin{theorem}\label{th:convKLDiv}
Let $0 < \tau < 2/3$ and let $d$ be the average degree of $G$. If $d \leq n^\tau$ then, recovery with high probability is possible only if
%\ba 
%\frac{m}{n} \geq \frac{1 - 3 \tau / 2}{2 \KLdiv{1/2}{\varepsilon}} \log \! \left( n \right) + o \! \left( \frac{\log n}{\KLdiv{1/2}{\varepsilon}} \right).\label{eq:convGenEps}
%\ea
\ba 
\frac{d}{\log n} \geq \frac{1 - 3 \tau / 2}{ \KLdiv{1/2}{\varepsilon}} - \frac{1}{\log n}  + o \! \left( \frac{1}{\KLdiv{1/2}{\varepsilon}} \right).\label{eq:convGenEps}
\ea
If $\varepsilon \rightarrow 1/2$, this condition implies
%\ba 
%\frac{m}{n} \geq \frac{1 - 3 \tau / 2}{\left( 1 - 2 \varepsilon\right)^2} \log \! \left( n \right) + o \! \left( \frac{\log n}{\left( 1 - 2 \varepsilon\right)^2} \right). \label{eq:convBigEps}
%\ea
\ba 
\frac{d}{\log n} \geq 2\frac{1 - 3 \tau / 2}{\left( 1 - 2 \varepsilon\right)^2} + o \! \left( \frac{1}{\left( 1 - 2 \varepsilon\right)^2} \right). \label{eq:convBigEps}
\ea
\end{theorem}

Before proving this Theorem, we compare it with the necessary condition 
\(
d \geq 2/(1 - H \! \left( \varepsilon \right) / \log 2), \label{eq:singer1}
\)
previously shown in \cite[Section~5]{ASinger_2011_angsync}. If $\varepsilon \in \left( 0,1/2 \right)$ does not depend on $n$, then this condition only implies $d = \Omega \! \left( 1 \right)$ and is thus weaker than $d = \Omega \! \left( \log n \right)$, which follows from Theorem~\ref{th:convKLDiv}. If $\varepsilon \to 1/2$, then $H \! \left( \varepsilon \right) = \log 2 - (1-2\eps)^2 / 2 + o \! \left( (1-2 \eps)^2 \right)$, and we can write the condition in \cite{ASinger_2011_angsync} as $1 - 2\eps = \Omega \bigl( \sqrt {1/d} \bigr)$. If there is a $\tau^\prime < 2/3$ for which $1 - 2\eps \geq n^{-\tau^\prime/2}$, then Theorem~\ref{th:convKLDiv} is tighter: it implies $1 - 2\eps = \Omega \bigl( \sqrt {\log \! \left( n \right)/d} \bigr)$. However, if there is no such $\tau^\prime$, then Theorem~\ref{th:convKLDiv} cannot be applied.\footnote{Using Slud's inequality \cite{slud77} to lower-bound $\Proba{\setE_j}$, one can improve the bound for $\varepsilon \rightarrow 1/2$ and show that whenever there is a $0 < \tau^\prime < 1$ for which $1 - 2\eps \geq n^{-\tau^\prime/2}$, then a necessary condition is $1 - 2\eps = \Omega \bigl( \sqrt {\log \! \left( n \right)/d} \bigr)$.}

\proof{[of Theorem~\ref{th:convKLDiv}] 
Fix a vertex $v_j$, and let $\setE_j$ denote the event that the variables attached to at least half of the edges that are incident to vertex~$v_j$ are noisy. As we argue next, if event $\setE_j$ occurs, then ML decoding recovers vertex-variables other than $x^V$ or $x^V \oplus 1^V$ with probability at least $1/2$. Indeed, if ML decoding correctly recovers the vertex-variables that are attached to the vertices adjacent to $v_j$ up to a global additive offset $\phi \in \left\{ 0,1 \right\}$, then---by assumption that event $\setE_j$ occurs---the probability that ML decoding recovers $x_j$ with offset $\phi \oplus 1$ is at least $1/2$. In particular, this implies that ML decoding can only be successful if the event $\bigcap_{v_j \in V} \setE_j^c$ occurs. Let $\setQ$ be an independent subset of $[n]$, i.e. a set such that no two vertices in it are adjacent. Since the noise $Z^E$ is drawn IID, the events $\set{\setE_j}_{j \in \setQ}$ are independent and the probability of the event $\bigcap_{j \in \setQ} \setE_j^c$ is easily computable. Moreover, the event $\bigcap_{j \in [n]} \setE_j^c$ can only occur if $\bigcap_{j \in \setQ} \setE_j^c$ occurs. A necessary condition for exact recovery thus is that the probability of the event $\bigcap_{j \in \setQ} \setE_j^c$ converges to one as the number of vertices increases. In the following, we prove the claim by identifying an independent set $\setQ$ and by upper-bounding the probability of the event $\bigcap_{j \in \setQ} \setE_j^c$.

Let $\text{deg} \! \left( v_j \right)$ be the degree of vertex~$v_j$, and assume w.l.o.g.\ $\text{deg} \! \left( v_1 \right) \leq \text{deg} \! \left( v_2 \right) \leq \ldots \leq \text{deg} \! \left( v_n \right)$. For every $0 < \delta \leq 1$
\be
d n \geq \sum^n_{j = \left\lceil \delta n \right\rceil} \text{deg} \! \left( v_j \right) \geq \left\lceil \left( 1 - \delta \right) n \right\rceil \text{deg} \! \left( v_{\left\lceil \delta n \right\rceil} \right).
\ee
For $j \leq \left\lceil \delta n \right\rceil$, we therefore find
\be 
\text{deg} \! \left( v_{j} \right) \leq \text{deg} \! \left( v_{\left\lceil \delta n \right\rceil} \right) \leq \frac{d n}{\left\lceil \left( 1 - \delta \right) n \right\rceil} \leq \frac{d}{1 - \delta}. \label{eq:ubDeg}
\ee
This implies that for every set $\setL \subseteq \set{ 1, \ldots, \left\lceil \delta n \right\rceil }$, the vertices $\set{v_j \colon j \in \setL}$ are disconnected from at least
\be\label{newlabel10}
\left\lceil \delta n \right\rceil - \card \setL \left( 1 + \frac{d}{ 1 - \delta } \right)
\ee
vertices in the set $\set{v_j \colon j \leq \left\lceil \delta n \right\rceil}$. We can construct an independent set $\setQ \subseteq \set{v_j \colon j \leq \left\lceil \delta n \right\rceil}$ by iteratively including vertices in $\setQ$ while keeping independence, until no vertex can be added. In fact, using the degree bound in (\ref{newlabel10}), it is easy to see that this process constructs an independent set $\setQ$ such that
\be
\card \setQ \geq \frac{\left\lceil \delta n \right\rceil}{1+\frac{d}{1 - \delta}} \geq \frac{\delta \left( 1 - \delta \right) n}{d + 1 - \delta}. \label{eq:lbCard}
\ee
To simplify notation, we introduce the variables $$a_j = \left\lfloor \frac{\text{deg} \! \left( v_j \right)}{2} \right\rfloor, \quad b_j = \left\lceil \frac{\text{deg} \! \left( v_j \right)}{2} \right\rceil.$$ If $j \leq \left\lceil \delta n \right\rceil$, then
\ba
\Proba{\setE_j} &= \sum^{\text{deg} \! \left( v_j \right)}_{k = b_j} \begin{pmatrix} \text{deg} \! \left( v_j \right) \\ k \end{pmatrix} \varepsilon^k \left( 1 - \varepsilon \right)^{\text{deg} \! \left( v_j \right) - k} \nonumber \\
&\geq \begin{pmatrix} \text{deg} \! \left( v_j \right) \\ b_j \end{pmatrix} \varepsilon^{b_j} \left( 1 - \varepsilon \right)^{a_j} \nonumber \\
&\stackrel{a)}\geq \frac{\sqrt{2 \pi \text{deg} \! \left( v_j \right)} \text{deg} \! \left( v_j \right)^{\text{deg} \! \left( v_j \right)} \varepsilon^{b_j} \left( 1 - \varepsilon \right)^{a_j}}{e^2 \sqrt{b_j a_j} b_j^{b_j} a_j^{a_j}} \nonumber \\
&\stackrel{b)}\geq \frac{2^{\text{deg} \! \left( v_j \right)}}{2 \sqrt{\text{deg} \! \left( v_j \right)}} \sqrt{\frac{\varepsilon}{1 - \varepsilon}} \varepsilon^{\frac{\text{deg} \! \left( v_j \right)}{2}} \left( 1 - \varepsilon \right)^{\frac{\text{deg} \! \left( v_j \right)}{2}} \nonumber \\
&= e^{-\frac{1}{2} \log \! \left( \frac{1 - \varepsilon}{\varepsilon} \right) - \log 2 - \text{deg} \! \left( v_j \right) \KLdiv{1/2}{\varepsilon} - \frac{1}{2} \log \! \left( \text{deg} \! \left( v_j \right) \right)} \nonumber \\
&\stackrel{c)}\geq e^{-\frac{1}{2} \log \! \left( \frac{1 - \varepsilon}{\varepsilon} \frac{d}{1 - \delta} \right) - \log 2 - \frac{d \KLdiv{1/2}{\varepsilon}}{1 - \delta}}, \label{eq:pjlb}
\ea
where $a)$ is due to Stirling's formula
\be 
1 \leq \frac{\ell !}{\sqrt{2 \pi \ell} \left( \ell /e \right)^\ell} \leq \frac{e}{\sqrt{2 \pi}}, \, \ell \in \naturals,
\een
$b)$ is due to the inequality of arithmetic and geometric means, the relation $\varepsilon / \left( 1-\varepsilon \right) < 1$, the fact that for every $t \geq 1$
\be
\frac{\left( t + \frac{1}{2} \right)^{t + \frac{1}{2}} \left( t - \frac{1}{2} \right)^{t - \frac{1}{2}}}{t^{2t}} = \left( 1 - \frac{1}{4 t^2} \right)^{t} \sqrt{\frac{1 + \frac{1}{2 t}}{1 - \frac{1}{2 t}}} < 1.3,
\een
and the inequality $2 \sqrt{2 \pi} / \left( 1.3 e^2 \right) \geq \frac{1}{2}$, and $c)$ is due to \eqref{eq:ubDeg}.

Since the events $\left\{ \setE^c_j \colon j \in \setQ \right\}$ are jointly independent,
\ba
\Proba{\bigcap_{j \in \setQ} \setE_j^c} &= \prod_{j \in \setQ} \left( 1 - \Proba{\setE_j} \right) \nonumber \\
&\stackrel{a)}{\leq} e^{-\sum_{j \in \setQ} e^{-\frac{1}{2} \log \! \left( \frac{1 - \varepsilon}{\varepsilon} \frac{d}{1 - \delta} \right) - \log \! \left( 2 \right) - \frac{d \KLdiv{1/2}{\varepsilon}}{1 - \delta}}} \nonumber \\
&\stackrel{b)}{\leq} e^{-e^{ \log \! \left( \frac{\delta \left( 1 - \delta \right) n}{2 (d + 1 - \delta)} \sqrt {\frac{1 - \delta}{d}} \sqrt{\frac{\varepsilon}{1-\varepsilon}} \right) - \frac{d \KLdiv{1/2}{\varepsilon}}{1 - \delta}}}, \label{eq:genCondmKLdiv}
\ea
where $a)$ holds since $1 - x \leq e^{-x}$ for $x \geq 0$ and because of \eqref{eq:pjlb}, and $b)$ is due to \eqref{eq:lbCard}. Clearly, a necessary condition for the RHS of \eqref{eq:genCondmKLdiv} to converge to $1$ is
\be 
\frac{d \KLdiv{1/2}{\varepsilon}}{1 - \delta} \geq \log \! \left( \frac{\delta \left( 1 - \delta \right) n}{2 (d + 1 - \delta)} \sqrt {\frac{1 - \delta}{d}} \sqrt{\frac{\varepsilon}{1-\varepsilon}} \right). \label{eq:genCondmKLdiv2}
\ee
Take $\delta = 1/ \log(n)$. Clearly, the average degree $d$ must be nonnegative. If $d \leq 1$, then
\ba 
&\log \! \left( \frac{\delta \left( 1 - \delta \right) n}{2 (d + 1 - \delta)} \sqrt {\frac{1 - \delta}{d}} \sqrt{\frac{\varepsilon}{1-\varepsilon}} \right) \nonumber \\
&\quad \geq \log n + \log \Biggl(\frac{\delta (1-\delta)^{\frac{3}{2}}}{2 \left( 2-\delta \right)}\Biggr) - \frac{1}{2} \log \biggl(\frac{1-\varepsilon}{\varepsilon}\biggr) \nonumber \\
&\quad \stackrel{(a)}\geq \log n + \log \Biggl(\frac{\delta (1-\delta)^{\frac{3}{2}}}{2 \left( 2-\delta \right)}\Biggr) - \frac{1}{2} \log \biggl(\frac{1}{\varepsilon (1-\varepsilon)}\biggr) \nonumber \\
&\quad \stackrel{(b)}\geq \log n + \Theta (\log \log n) - \KLdiv{1/2}{\varepsilon}, \label{eq:genCondmKLdiv3}
\ea
where $(a)$ is due to $1 - \varepsilon \leq 1$, and $(b)$ holds because $\delta = 1 / \log n$ and since $\KLdiv{1/2}{\varepsilon} = - \log 2 - \log ( \varepsilon (1-\varepsilon)) / 2$. If $1 < d \leq n^\tau$, then
\ba 
&\log \! \left( \frac{\delta \left( 1 - \delta \right) n}{2 (d + 1 - \delta)} \sqrt {\frac{1 - \delta}{d}} \sqrt{\frac{\varepsilon}{1-\varepsilon}} \right) \nonumber \\
&\quad = \log \! \left(n d^{-\frac{3}{2}} \right) + \log \Biggl(\frac{\delta (1-\delta)^{\frac{3}{2}}}{2 \left( 1+\frac{1-\delta}{d} \right)}\Biggr) + \frac{1}{2} \log \biggl(\frac{\varepsilon}{1-\varepsilon}\biggr) \nonumber \\
&\quad \stackrel{(a)}\geq \left( 1 - \frac{3 \tau}{2} \right) \log n - \KLdiv{1/2}{\varepsilon} + \Theta (\log \log n), \label{eq:genCondmKLdiv4}
\ea
where $(a)$ holds since $d \leq n^\tau$, because $\delta = 1 / \log(n)$, since $1 - \varepsilon \leq 1$, and because $\KLdiv{1/2}{\varepsilon} = - \log 2 - \log ( \varepsilon (1-\varepsilon)) / 2$. For $d \leq n^\tau$, we thus obtain from \eqref{eq:genCondmKLdiv3} (if $d \leq 1$) or \eqref{eq:genCondmKLdiv4} (if $d > 1$) that \eqref{eq:genCondmKLdiv2} cannot hold unless \eqref{eq:convGenEps} holds.
}

\subsection{Sufficient Conditions for Successful Recovery}

We next present sufficient conditions for exact recovery. We first focus on graphs from the Erd\H{o}s-R\'enyi model. Then, we consider arbitrary graphs and present a condition that is sufficient for every graph and depends only on the graph's Cheeger constant.\\

For a random base-graph $G = (V,E)$ from the Erd\H{o}s-R\'enyi model, we require the vertex-variables $x^V$ to be recoverable from the edge-variables $Y^E$ except with some probability that vanishes as the number of vertices increases.

\begin{theorem}\label{th:suffCondErdRen}\label{IT_sufficient_ErdosReinyi}
Suppose the base-graph is drawn from the Erd\H{o}s-R\'enyi model $ER \! \left( n, p \right)$ with $p > 2 \log n / n$, and let $d$ denote its expected average degree, i.e., $d = (n-1)p$. Then the condition
\be
\frac{d}{\log n} \geq \frac{1}{\left( 1-\sqrt{\frac{2 \log n}{d}} \right) \KLdiv{1/2}{\varepsilon}}  + o \! \left( \frac{1}{\KLdiv{1/2}{\varepsilon}} \right) \label{eq:suffCondRandGraphThm}
\ee
%\be
%\frac{m}{n} \geq \frac{1}{2 \left( 1-\sqrt{\frac{\log \! \left( n \right)}{m/n}} \right) \KLdiv{1/2}{\varepsilon}} \log \! \left( n \right) + o \! \left( \frac{\log \! \left( n \right)}{\KLdiv{1/2}{\varepsilon}} \right) \label{eq:suffCondRandGraphThm}
%\ee
is sufficient to guarantee exact recovery with high probability. If $\epsilon \rightarrow 1/2$, the condition is
\be 
\frac{d}{\log n} \geq \frac{2}{\left( 1 - 2 \epsilon \right)^2}  + o \! \left( \frac{1}{\left( 1 - 2 \varepsilon \right)^2} \right). \label{eq:suffCondErdRenBigEps}
\ee
%\be 
%\frac{m}{n} \geq \frac{1}{\left( 1 - 2 \epsilon \right)^2} \log \! \left( n \right) + o \! \left( \frac{\log \! \left( n \right)}{\left( 1 - 2 \varepsilon \right)^2} \right). \label{eq:suffCondErdRenBigEps}
%\ee
\end{theorem}

\proof{
Let $x^V$ be the vertex-variables, and denote by $d_H \! \left( \cdot, \cdot \right)$ the Hamming distance. ML decoding recovers the vertex-variables $x^V$ from the measurements $Y^E = B_G x^V \oplus Z^E$ if every binary $n$-tuple $\tilde x^V \notin \left\{ x^V, x^V \oplus 1^V \right\}$ satisfies
\be 
d_H \! \left( Y^E, B_G \tilde x^V \right) > d_H \! \left( Y^E, B_G x^V \right). \label{eq:MLRecCond}
\ee
Since $d_H \! \left( x^V, \tilde x^V \oplus 1^V \right) = n - d_H \! \left( x^V, \tilde x^V \right)$ and $B_G \tilde x^V = B_G \left( \tilde x^V \oplus 1^V \right)$, assume w.l.o.g.\ $d_H \! \left( x^V, \tilde x^V \right) \leq \left\lfloor n/2 \right\rfloor$. For $x^V \in \left\{ 0,1 \right\}^n$ let $\setD_{x^V} \subseteq \left\{ 0,1 \right\}^m$ contain all vectors $y^E \in \left\{ 0,1 \right\}^m$ for which ML decoding recovers $x^V$ or $x^V \oplus 1^V$, i.e., $y^E \in \setD_{x^V}$ iff \eqref{eq:MLRecCond} holds for all binary $n$-tuples $\tilde x^V$ satisfying $1 \leq d_H \! \left( x^V, \tilde x^V \right) \leq \left\lfloor n/2 \right\rfloor$. Since the mapping $x^V \mapsto B_G x^V$ is linear, we find $\setD_{x^V} = \setD_{0^V} \oplus B_G x^V$ and
\be 
\Proba{Y^E \notin \setD_{x^V}} = \Proba{Z^E \notin \setD_{0^V}}. \label{eq:linCode}
\ee
We thus assume w.l.o.g.\ $x^V = 0^V$. Let $\tilde x^V$ be a binary $n$-tuple that satisfies $1 \leq d_H \! \left( 0^V, \tilde x^V \right) \leq \left\lfloor n/2 \right\rfloor$, and suppose the ML decoder has to decide between the two hypotheses $0^V$ and $\tilde x^V$. Clearly, it decodes $\tilde x^V$ only if $d_H \! \left( Z^E, B_G \tilde x^V \right) \leq d_H \! \left( Z^E, B_G 0^V \right)$. If we let $\setT = \set{ i \colon \elementof{B_G \tilde x^V}{i} = 1}$ be the set of edges $e_i$ such that $x_{i_1} \oplus x_{i_2} \neq \tilde x_{i_1} \oplus \tilde x_{i_2}$, then this implies that the ML decoder decides for $\tilde x^V$ only if at least half of the edge-variables $\left\{ Y_i \right\}_{i \in \setT}$ are corrupted, i.e.,
\ba
\sum_{i \in \setT} Z_i \geq \card \setT/2. \label{eq:pairwise0}
\ea
The Chernoff-H\"{o}ffding theorem implies
\ba
&\Proba{\sum_{i \in \setT} \left( Z_i - \varepsilon \right) \geq \card \setT \left( 1/2 - \varepsilon \right)} \nonumber \\
&\quad \leq e^{-D \! \left( 1/2 || \varepsilon \right) \card \setT } \label{eq:pairwise}.
\ea
Moreover, the cardinality of the set $\setT$ is nothing else but the cut of the set of vertices $v_i$ for which $x_i$ and $\tilde x_i$ are distinct in the sense that $x_i = 0$ and $\tilde x_i = 1$, i.e., for $\setS = \set{v_j \colon \tilde x_j = 1}$ it holds that $\card \setT = \text{cut} \! \left( \setS \right)$. Take $\delta > 0$, and let $\setE$ be the event that $\text{cut} \! \left( \setS \right) > \left( 1 - \delta \right) p \, \card \setS \left( n - \card{\setS} \right)$ holds for all subsets $\setS$ of $V$. Since the graph is from the Erd\H{o}s-R\'enyi model $ER \! \left( n, p \right)$, we find for every $\nu, \eta > 0$,
\ba
\Proba{\setE^c} &=  \Proba{\exists \, \setS \subseteq V \colon \text{cut} \! \left( S \right) \leq \left( 1 - \delta \right) \card \setS \left( n - \card \setS \right) p} \nonumber \\
& \leq \sum^{\left\lfloor \frac{n}{2} \right\rfloor}_{k = 1} \sum_{\setS \colon \card \setS = k} \Proba{\text{cut} \! \left( S \right) \leq \left( 1 - \delta \right) k \left( n - k \right) p} \nonumber 
\ea
hence 
\ba
\Proba{\setE^c} &\stackrel{(a)} \leq \sum^{\left\lfloor \nu n \right\rfloor}_{k = 1} \begin{pmatrix} n \\ k \end{pmatrix} e^{ - \left( \delta + \left( 1 - \delta \right) \log \! \left( 1 - \delta \right) \right) k \left( n - k \right) p } + \nonumber\\
 &\qquad + \sum^{\left\lfloor \frac{n}{2} \right\rfloor}_{k = \left\lfloor \nu n \right\rfloor+1} \begin{pmatrix} n \\ k \end{pmatrix} e^{ - \left( \delta + \left( 1 - \delta \right) \log \! \left( 1 - \delta \right) \right) k \left( n - k \right) p } \nonumber \\
&\stackrel{(b)}\leq \sum^{\left\lfloor \nu n \right\rfloor}_{k = 1}  e^{ - k \left( \left( \delta + \left( 1 - \delta \right) \log \! \left( 1 - \delta \right) \right) \left( 1 - \frac{k}{n} \right) n p - \log n \right) } + \nonumber \\
 &\qquad + \sum^{\left\lfloor \frac{n}{2} \right\rfloor}_{k = \left\lfloor \nu n \right\rfloor+1} e^{- n \left( \left( \delta + \left( 1 - \delta \right) \log \! \left( 1 - \delta \right) \right) \frac{k}{n} \left( 1 - \frac{k}{n} \right) n p - H \! \left( \frac{k}{n} \right) - \eta \right)} \nonumber \\
& \stackrel{(c)}\leq \frac{e^{ - \left( \left( \delta + \left( 1 - \delta \right) \log \! \left( 1 - \delta \right) \right) \left( 1 - \nu \right) d - \log n \right) }}{1 - e^{ - \left( \left( \delta + \left( 1 - \delta \right) \log \! \left( 1 - \delta \right) \right) \left( 1 - \nu \right) d - \log n \right) }} \nonumber \\
 &\qquad + e^{- n \left( \nu \left( 1 - \nu \right) \left( \delta + \left( 1 - \delta \right) \log \! \left( 1 - \delta \right) \right) d - \log 2 - \eta - \frac{\log n }{n}\right)}, \label{eq:cutBound}
\ea
where $(a)$ is due to the multiplicative Chernoff bound, $(b)$ holds since for $n$ large ${n \choose k}$ is upper-bounded by $n^k$ as well as $e^{n \left( H \! \left( k/n \right) + \eta \right)}$, where $H ( k/n ) = k/n \log (n/k) + (1-k/n) \log (n/(n-k))$, and $(c)$ is true because $d = \left( n - 1 \right) p$, binary entropy satisfies $H (k/n) \leq \log 2$, and $a \left( 1 - a \right)$ is concave on $\left[0,1\right]$. Moreover, the union bound implies for every $\nu, \eta > 0$ and sufficiently large $n$
\ba
\Proba{Y^E \notin \setD_{x^V} \left| \setE \right.} %\nonumber \\
%&\quad
 &\leq \sum_{\tilde x^V} \Proba{\left. \sum_{i \in \setT} Z_i \geq \card \setT / 2 \right| \setE } \nonumber \\
& \leq \sum^{\left\lfloor \nu n \right\rfloor}_{k = 1} \begin{pmatrix} n \\ k \end{pmatrix} e^{ -D \! \left( 1/2 || \varepsilon \right) \left( 1 - \delta \right) k \left( n - k \right) p } + \nonumber \\
& \quad \quad + \sum^{\left\lfloor \frac{n}{2} \right\rfloor}_{k = \left\lfloor \nu n \right\rfloor + 1} \begin{pmatrix} n \\ k \end{pmatrix} e^{ -D \! \left( 1/2 || \varepsilon \right) \left( 1 - \delta \right) k \left( n - k \right) p } \nonumber \\
&\leq \sum^{\left\lfloor \nu n \right\rfloor}_{k = 1} e^{ -k \left( D \! \left( 1/2 || \varepsilon \right) \left( 1 - \delta \right) \left( 1 - \frac{k}{n} \right) n p - \log n \right) } + \nonumber \\
& \quad \quad + \sum^{\left\lfloor \frac{n}{2} \right\rfloor}_{k = \left\lfloor \nu n \right\rfloor + 1} e^{ -n \left( D \! \left( 1/2 || \varepsilon \right) \left( 1 - \delta \right) \frac{k}{n} \left( 1 - \frac{k}{n} \right) n p - H \! \left( \frac{k}{n} \right) - \eta \right) } \nonumber \\
& \leq \frac{e^{ -\left( \left( 1 - \delta \right) \left( 1 - \nu \right) D \! \left( 1/2 || \varepsilon \right) d - \log n \right) }}{1 - e^{ -\left( \left( 1 - \delta \right) \left( 1 - \nu \right) D \! \left( 1/2 || \varepsilon \right) d - \log n \right) }} + \nonumber \\
& \quad \quad + e^{- n \left( \left( 1 - \delta \right) \nu \left( 1 - \nu \right) D \! \left( 1/2 || \varepsilon \right) d - \log 2 - \eta - \frac{\log n}{n} \right)}. \label{eq:pairwiseBound}
\ea
The law of total probability implies that
\ba
\!\! \Proba{Y^E \notin \setD_{x^V}} \leq \Proba{Z^E \notin \setD_{0^V} \left| \setE \right.} + \Proba{\setE^c}. \label{eq:boundFalseDecZero}
\ea
From \eqref{eq:linCode} and \eqref{eq:pairwiseBound}--\eqref{eq:boundFalseDecZero} we conclude that ML decoding succeeds if $\left( \log 2 + \eta \right)/\nu < \log n$ and
\ba
d &> \frac{1}{\left( \delta + \left( 1 - \delta \right) \log \! \left( 1 - \delta \right) \right) \left( 1 - \nu \right)} \log n \\
d &> \frac{1}{D \! \left( 1/2 \left| \left| \varepsilon \right. \right. \right) \left( 1 - \delta \right) \left( 1 - \nu \right)} \log n.
\ea
If we choose $\eta = 1$ and $\nu = o \! \left( 1 \right)$ so that $1/\nu = o \! \left( \log n \right)$, then we find that the following conditions are sufficient
\ba
d &> \frac{1}{\left( \delta + \left( 1 - \delta \right) \log \! \left( 1 - \delta \right) \right)} \left( \log n + o \! \left( \log n \right) \right) \\
d &> \frac{1}{D \! \left( 1/2 \left| \left| \varepsilon \right. \right. \right) \left( 1 - \delta \right)} \left( \log n + o \! \left( \log n \right) \right).
\ea
Since $\delta^2 / 2 \leq \delta + \left( 1 - \delta \right) \log \! \left( 1 - \delta \right)$ for $\delta \in \left( 0,1 \right)$, the above two constraints are satisfied if \eqref{eq:suffCondRandGraphThm} holds.
}\\

In the proof of Theorem~\ref{th:suffCondErdRen}, we used the fact that, for a graph from the Erd\H{o}s-R\'enyi model $ER \! \left( n, p \right)$, the cut of each subset $\setS \subseteq V$ is with high probability approximately as large as its expectation, i.e., for $\delta > 0$ it holds with high probability that
\ba
\text{cut} \! \left( \setS \right) > \left( 1 - \delta \right) p \, \card \setS \left( n - \card{\setS} \right), \, \forall \, \setS \subseteq V. \label{eq:motForCheeger}
\ea
For every set $\setS \subseteq V$, define
\ba
\text{vol} \! \left( \setS \right) = \sum_{v \in \setS} \text{deg} \! \left( v \right).
\ea
Note that $\Ex{\text{vol} \! \left( \setS \right)} = p \, \card \setS \left( n - 1 \right)$ and $\Ex{\text{cut} \! \left( \setS \right)} = p \, \card \setS \left( n - \card \setS \right)$. Moreover, the multiplicative Chernoff bound implies that $\text{vol} \! \left( \setS \right) \leq \left( 1 + \delta \right) p \, \card \setS \left( n - 1 \right)$ holds with high probability. Hence, instead of \eqref{eq:motForCheeger} we could require that for some $\mu \in \left( 0,1 \right)$ and for every $\setS \subseteq V$ with $\card \setS \leq n - \card \setS$

\be
\frac{\text{cut} \! \left( \setS \right)}{\text{vol} \! \left( \setS \right)} > \left( 1 - \mu \right) \frac{n - \card \setS}{n - 1}. \label{eq:condCutVol}
\ee
Recalling that the Cheeger constant $h_G$ of a graph is
\begin{equation}\label{def:Cheegerconstant44}
h_G = \min_{\setS \subseteq \left[n\right]} \frac{\text{cut} \! \left( \setS \right)}{\min \! \left\{ \text{vol} \! \left( \setS \right), \text{vol} \! \left( \setS^c \right)\right\}},
\end{equation}
it is clear that \eqref{eq:condCutVol} holds for every subset $\setS \subseteq V$ if
\be 
h_G > \left( 1 - \mu \right) \frac{1}{2}.
\een

This motivates our next result, which is a recovery guarantee in terms of the Cheeger constant:

\begin{theorem}\label{th:CheegerDirPart}\label{IT_sufficient_CheegerConstant}
If the base-graph $G = \left( V,E \right)$ has Cheeger constant $h_G$ and the minimum degree satisfies
\be 
\frac{\min_j \text{deg} \! \left( v_j \right)}{\log n} > \frac{1}{h_G \KLdiv{1/2}{\varepsilon}}, \label{eq:cheegSuffCond}
\ee
then exact recovery with high probability is possible. In particular, if the base-graph $G = \left( V,E \right)$ is $d$-regular, then a sufficient condition for exact recovery is
\be 
\frac{d}{\log n} > \frac{1}{ h_G \KLdiv{1/2}{\varepsilon}}. \label{eq:cheegConstSuffCond}
\ee
% \eqref{eq:cheegSuffCond} is equivalent to
% \be 
% c > \frac{2}{h_G \left( 1 - 2 \varepsilon\right)^2} + o \! \left( \frac{1}{h_G \left( 1 - 2 \varepsilon\right)^2} \right),
% \ee
% and
If $\epsilon \rightarrow 1/2$, then \eqref{eq:cheegConstSuffCond} is equivalent to
\be 
\frac{d}{\log n} > \frac{2}{h_G \left( 1 - 2 \varepsilon\right)^2} + o \! \left( \frac{1}{h_G \left( 1 - 2 \varepsilon\right)^2} \right).
\label{eq:cheegSuffCond_23}
\ee
\end{theorem}

\proof{
Denote $c = \min_j \text{deg} \! \left( v_j \right) / \log n$. Because of \eqref{eq:linCode}--\eqref{eq:pairwise}, the union bound, and since $\card \setT = \text{cut} \! \left( \setS \right) \geq h_G \, \text{vol} \! \left( \setS \right) \geq c \, h_G \, \card{\setS} \log \! \left( n \right)$ holds for every subset $\setS \subseteq V$ with $\card \setS \leq n/2$, we find that
\ba
&\Proba{Y^E \notin \setD_{x^V}} = \Proba{Z^E \notin \setD_{0^V}} \nonumber \\
&\quad \leq \frac{1}{2} \sum_{\tilde x^V \notin \left\{ 0^V, 1^V \right\} } \Proba{d_H \! \left( Z^E, B_G x^V \right)  \leq d_H \! \left( Z^E, B_G 0^V \right)} \nonumber \\
&\quad \leq \sum^{\left\lfloor \frac{n}{2} \right\rfloor}_{k = 1} \begin{pmatrix} n \\ k \end{pmatrix} e^{ - k c \, h_G D \! \left( 1/2 || \varepsilon \right) \log n } \nonumber \\
&\quad \leq \sum^{\left\lfloor \frac{n}{2} \right\rfloor}_{k = 1} e^{ -k \left( c \, h_G D \! \left( 1/2 || \varepsilon \right) \log n - \log n \right) } \nonumber \\
&\quad \leq \frac{e^{- \left( c \, h_G D \! \left( 1/2 || \varepsilon\right) \log n - \log n \right)}}{1 - e^{ - \left( c \, h_G D \! \left( 1/2 || \varepsilon \right) \log n - \log n \right)}}. \label{eq:boundCheeger}
\ea
Hence, if \eqref{eq:cheegSuffCond} holds, then ML decoding recovers the correct vertex-variables $x^V$.
}\\

Interestingly, if the base-graph is drawn from the Erd\H{o}s-R\'enyi model $\mathrm{ER} \! \left( n, p \right)$, then the sufficient conditions of Theorem~\ref{th:suffCondErdRen} and Theorem~\ref{th:CheegerDirPart} exhibit the same scaling behavior:

\begin{remark}
If the base-graph is drawn from the Erd\H{o}s-R\'enyi model $ER \! \left( n, p \right)$, then it has a non-vanishing spectral gap for $p > C \log n / n$ (see~\cite{jiang12}). Moreover, for every $\delta \in \left( 0,1 \right)$ and $p > 2 \log n / \left( \delta^2 n \right)$
\be
\Proba{\exists \, \setS \colon \text{vol} \! \left( \setS \right) \leq \left( 1 - \delta \right) p \, \card \setS \left( n - 1 \right)} \rightarrow 0 \left( n \rightarrow \infty \right).
\een
Observe that if $\text{vol} \! \left( \setS \right) > \left( 1 - \delta \right) p \, \card \setS \left( n - 1 \right)$ for every $\setS \subseteq V$, then $\min_{j} \text{deg} \! \left( v_j \right) \geq \left( 1 - \delta \right) \left( n - 1 \right) p$.
\end{remark}

It is natural to give recovery guarantees in terms of the Cheeger constant: A graph with a small minimum cut consists of two rather disconnected components so that the probability of decoding one component without additive offset and the other component with constant additive offset $1$ is non-negligible. As we argue next, deriving a necessary condition that bounds the Cheeger constant away from zero is, however, impossible. Indeed, suppose the base-graph consists of two equally sized components, which are connected by $\log n$ edges. Moreover, assume the two graphs that are obtained by disconnecting the two components have Cheeger constant $h_G$ and minimum degree $c \log n$, where $c$ is some positive constant for which the sufficient condition \eqref{eq:cheegConstSuffCond} of Theorem~\ref{th:CheegerDirPart} holds. Then, Theorem~\ref{th:CheegerDirPart} implies that each component can be recovered correctly (up to an inevitable additive offset). Moreover, with high probability less than half of the $\log n$ edges that connect the two components are corrupted by noise. Hence, ML decoding indeed recovers the correct vertex-variables up to a constant additive binary offset. But the Cheeger constant of the graph satisfies $h_g \leq 2 / \left( c n \right)$ and thus converges to zero as $n$ approaches infinity.
This leaves the interesting open question of investigating a characteristic of the graph that captures how easy it is to solve (on it) the type of inverse problems considered here.

%%%%%%%%%%%%%%%%%%%%%%%%%%%%%%%%%%%%%%%%%%%%%%%%%%%%%%%%%%%%%%%%%%%%%%%%%%

\section{Computationally efficient recovery - the SDP}

In this section we analyze a tractable method to recover $x^V$ from the noisy measurements $Y^E$, which is based on SDP. Ideally, one would like to find the maximum likelihood estimator $x^\ast = \argmin_{x_i \in \{0,1\}} \sum_{(i,j)\in E} 1_{\left\{x_i  \neq y_{(i,j)} \oplus x_j \right\}}$. By defining the $\{\pm1\}$-valued variables $g_i = (-1)^{x_i}$ and the coefficients $\rho_{ij}=(-1)^{y_{(i,j)}}$, the ML problem is reformulated as
\begin{equation}\label{angsynch:withgs}
\min_{g_i\in\{\pm1\}} \sum_{(i,j)\in E} (g_i - \rho_{ij}g_j)^2.
\end{equation}
This problem is known to be NP-hard in general (in fact, it is easy to see that it can encode \texttt{Max-Cut}). In what follows, we will describe and analyze a tractable algorithm, which was first proposed in~\cite{ASinger_2011_angsync} to approximate the solution of (\ref{angsynch:withgs}). We will state conditions under which the algorithm is able to recover the vertex-variables $x^V$. The idea is to consider a natural semidefinite relaxation. Other properties of this SDP have been studied in \cite{NAlon_ANaor_2006,Bandeira_LittleGrothendieckOd}.

Let $W$ be the $n\times n$ matrix with $W(i,j) = \rho_{ij}$ if $(i,j)\in E$ and $W(i,j) = 0$ otherwise. Problem (\ref{angsynch:withgs}) has the same solutions as
\(
\max_{g_i\in\{\pm1\}} \tr\left[W gg^T\right],
\)
which in turn is equivalent to
\begin{align}
\max \  &\tr\left[W X\right] \label{noSDP_Z2Synch}\\
\text{s.t.}\  & X\in\RR^{n\times n},\ X_{ii} = 1\ \forall i,\   X\succeq 0,\  \rank(X) =1.\notag
%& X_{ii} = 1,\ \forall i \notag\\
%\notag %\\
%& \rank(X) =1\notag.
\end{align}
(Given the optimal rank 1 solution $X$ of (\ref{noSDP_Z2Synch}), $g_i = (-1)^{x_{i}}$ is the only non-trivial eigenvector of $X$.) As the rank constraint is non-convex, we consider the following convex relaxation
\begin{equation}\label{SDP_Z2Synch}
 \max  \tr\left[WX\right] \quad \text{s.t. } X_{ii}=1, \, X  \succeq 0.
\end{equation}
%\begin{align}
%\min \  &\tr(W_1 X) \notag\\
%\text{s. t.}\  & X\in\RR^{n\times n} \label{SDP_Z2Synch}\\
%& X_{ii} = 1,\ \forall i \notag\\
%&  X\succeq 0\notag.
%\end{align}
Note that (\ref{SDP_Z2Synch}) is an SDP and can be solved, up to arbitrary precision, in polynomial time \cite{LVanderberghe_SBoyd_1996}. Note that a solution of (\ref{SDP_Z2Synch}) need not be rank 1 and thus need not be a solution of (\ref{noSDP_Z2Synch}). However, we will show that under certain conditions \eqref{SDP_Z2Synch} recovers the same optimal solution as \eqref{noSDP_Z2Synch}. In this case, $g_i = (-1)^{x_{i}}$ is the only non-trivial eigenvector of $X$ and $x^V$ can be recovered via the tractable program (\ref{SDP_Z2Synch}).

\textbf{Notation:} Recall that $G$ is the underlying graph on $n$ nodes, and let $H$ be the subgraph representing the incorrect edges (corresponding to $Z_{(i,j)}=1$). Let $A_G$, $A_H$, $D_G$, $D_H$, $L_G$, and $L_H$ be, respectively, the adjacency, degree, and Laplacian matrices of the graphs $G$ and $H$.

As in \cite{ASinger_2011_angsync}, we assume w.l.o.g.\footnote{It is not difficult to see that the recovery success of either (\ref{noSDP_Z2Synch}) or (\ref{SDP_Z2Synch}) only depends on which edges are correct and which are incorrect, and not on the values of $x^V$ (or $g$).}\ that $x^V \equiv 0$ so that $g \equiv 1$. Then, $W = A_G -2A_H$, and (\ref{SDP_Z2Synch}) can be rewritten as:
\begin{equation}\label{SDP_Z2Synch00}
 \max  \tr\left[(A_G -2A_H)X\right] \quad \text{s.t. } X_{ii}=1, \, X  \succeq 0.
\end{equation}

Our objective is to understand when $X=gg^T=11^T$ is the unique optimal solution to (\ref{SDP_Z2Synch00}). The dual of the SDP is
\begin{equation}\label{SDP_Z2Synch2dual}
 \min  \tr(Q) \quad \text{s.t. } Q\text{ diagonal}, \, Q-\left(A_G -2A_H\right)  \succeq 0.
\end{equation}
Duality guarantees that the objective value of (\ref{SDP_Z2Synch00}) cannot exceed that of (\ref{SDP_Z2Synch2dual}). Thus, if there exists $Q$, feasible solution of (\ref{SDP_Z2Synch2dual}), such that $\tr(Q) = \tr\left[\left(A_G -2A_H\right)11^T\right]$, then $X=11^T$ is an optimal solution of (\ref{SDP_Z2Synch00}). Moreover, $Q$ and $11^T$ have to satisfy complementary slackness: $\tr(11^T(Q-\left(A_G -2A_H\right)))=0$. Given these constraints, one can ask that the equality holds for each row partial sum and construct the natural candidate $Q=D_G - 2D_H$. Indeed, it is easy to see that $\tr(D_G-2D_H) = \tr\left[\left(A_G -2A_H\right)11^T\right]$. Hence, if
\begin{equation}\label{conditionneededdual0}
 L_G - 2L_H = D_G - 2D_H -\left(A_G -2A_H\right) \succeq 0,
\end{equation}
i.e., the dual variable is positive-semidefinite (PSD), then $11^T$ must be an optimal solution of (\ref{SDP_Z2Synch00}). Additionally, if $L_G - 2L_H$ is not only PSD but also its second smallest eigenvalue is non-zero, since the complementarity conditions guarantee that any optimal solution $X'$ needs to satisfy $\tr\left(X'(L_G - 2L_H)\right)=0$, it is not difficult to show that any optimal solution needs to be a multiple of $11^T$. As one can easily see from the constraints of the SDP that no other multiple of $11^T$ is a feasible solution, $11^T$ must be the unique optimal solution. Since the success of (\ref{SDP_Z2Synch00}) does not depend on the value $g_i = (-1)^{x_{i}}$ of the ground truth, we have thus shown:

\begin{lemma}
If 
\begin{equation}\label{conditionneededdual}
 L_G - 2L_H\succeq 0 \text{ and } \lambda_2(L_G - 2L_H) > 0,
\end{equation}
then $gg^T$, where $g_i = (-1)^{x_{i}}$ corresponds to the ground truth, is the unique solution to (\ref{SDP_Z2Synch}).
\end{lemma}

\subsection{Erd\H{o}s-R\'enyi Model}

We now assume that the underlying graph is drawn from the Erd\H{o}s-R\'enyi model $\mathrm{ER}(n,p)$ and use condition (\ref{conditionneededdual}) to give guarantees for exact recovery.

For each pair of vertices $i<j$, let $\Lambda_{ij}$ be an $n\times n$ symmetric  matrix with $\Lambda_{ij}(i,i) = 1$, $\Lambda_{ij}(j,j) = 1$, $\Lambda_{ij}(i,j) = \Lambda_{ij}(j,i) = -1$, and $\Lambda_{ij}(k,l) = 0$ for all other pairs $(k,l)$. Observe that $\Lambda_{ij}\succeq 0$, and
\(
 L_G = \sum_{i<j:(i,j)\in E} \Lambda_{ij}.
\)
Let $\alpha_{ij}$ be the random variable that takes the value $0$ if edge $(i,j)$ is not in $G$, the value $1$ if it is in $G$ but not in $H$, and the value $-1$ if it is in $H$. Hence $\alpha_{ij}$ are i.i.d.\ with distribution % and take the values $0,1,-1$ with probability, respectively, $1-p$, $p \left(1-\varepsilon\right)$, and $p \varepsilon$.
 \[
  \alpha_{ij} = \left\{ \begin{array}{cl}  
    0 & \text{ with probability } 1-p \\
     1 & \text{ with probability } p \left(1-\varepsilon\right) \\ 
     -1 & \text{ with probability } p \varepsilon.
       \end{array}   \right.
 \]

In the new notation,
\[
 L_G - 2L_H = \sum_{i< j} \alpha_{ij}\Lambda_{ij}.
\]
We define the centered random variables
\(
 A_{ij} = \left(p(1-2\varepsilon)-\alpha_{ij}\right)\Lambda_{ij}.
\)
For $A = \sum_{i< j} A_{ij}$, we can write
\[
 L_G - 2L_H =% p(1-2\varepsilon)\sum_{i\leq j}\Lambda_{ij} - \sum_{i\leq j} X_{ij} = 
 p(1-2\varepsilon)(nI-11^T) - A.
\]
Since $\Lambda_{ij}$ always contains the vector $1$ in the null-space, (\ref{conditionneededdual}) is equivalent to
\(
 \lambda_{\max}(A) < p(1-2\varepsilon)n.
\)
%where $\lambda_{\max}$ is the largest eigenvalue orthogonal to the subspace generated by $\1$.

We are now interested in understanding for which values of $p$, $\varepsilon$, and $n$ there is some $\delta>0$ such that
\[
 \Prob\left[ \lambda_{\max}(A) \geq p(1-2\varepsilon)n  \right] \leq n^{-\delta}.
\]
To this end, we use the Matrix Bernstein inequality (Theorem 1.4 in \cite{Tropp:TailBoundsRM}), which implies
\[
 \Prob\left[ \lambda_{\max}(A) \geq t  \right] \leq n \exp\left( - \frac{t^2/2}{\sigma^2+Rt/3} \right),
\]
where $\sigma^2 = \bigl\| \sum_{i<j} \EE A_{ij}^2 \bigr\|$, with $\|\cdot\|$ denoting the spectral norm, and $R\geq\lambda_{\max}\left(A_{ij}\right)$. Note that
\begin{align*}
 \sigma^2 & = \left\| \sum_{i<j} \EE A_{ij}^2 \right\| = \left\| \sum_{i<j} \EE(p(1-2\varepsilon)-\alpha_{ij})^22\Lambda_{ij} \right\| \\
 & = 2\EE(p(1-2\varepsilon)-\alpha_{ij})^2\left\| \sum_{i<j} \Lambda_{ij} \right\| \\ 
 &= 2n \EE(p(1-2\varepsilon)-\alpha_{ij})^2,
\end{align*}
which gives $\sigma^2 = 2np\left[1 - p(1-2\varepsilon)^2 \right]$.
%\(
 %\sigma^2 =  2np\left[1 - p(1-2\varepsilon)^2 \right]
%\sigma^2 =  2n \EE\left[p(1-2\varepsilon)-\alpha_{ij}\right]^2 = 2np\left[1 - p(1-2\varepsilon)^2 \right],
%\)
%
%and
%\[
% \EE(p(1-2\varepsilon)-\alpha_{ij})^2 = p\beta(p,\varepsilon),
%\]
%\[
% \EE(p(1-2\varepsilon)-\alpha_{ij})^2 = (1-p)(1-2\varepsilon)^2p^2 + p\left(1-\varepsilon\right)((1-2\varepsilon)p-1)^2 + p \varepsilon((1-2\varepsilon)p+1)^2 = p\beta(p,\varepsilon),
%\]
%where we define $\beta(p,\varepsilon)$ as
%\begin{equation}\label{def:betaqp}
% \beta(p,\varepsilon) = (1-p)(1-2\varepsilon)^2p + \left(\frac{1+(1-2\varepsilon)}2\right)((1-2\varepsilon)p-1)^2 + \left(\frac{1-(1-2\varepsilon)}2\right)((1-2\varepsilon)p+1)^2.
%\end{equation}
Also, 
\(
 \lambda_{\max}(A_{ij}) \leq 2p(1-2\varepsilon)+2.
\)
Setting $t = p(1-2\epsilon)n$ gives
\begin{align*}
& \Prob\left[ \lambda_{\max}(A) \geq p(1-2\varepsilon)n  \right]  \\ 
 %&\quad \leq n\exp\left( -\frac{(p(1-2\varepsilon)n)^2/2}{2np\left[1 - p(1-2\varepsilon)^2 \right]+(2p(1-2\varepsilon)+2)p(1-2\varepsilon)n/3} \right) \\ 
 &\quad \leq  n\exp\left( -\frac14\frac{(1-2\varepsilon)^2}{1 - \frac23p(1-2\varepsilon)^2+\frac13(1-2\varepsilon)} pn\right),
\end{align*}
which together with (\ref{conditionneededdual}) concludes the proof of the following Theorem:
\begin{theorem}\label{SDP_ErdosReinyi}
Let $d$ be the expected average degree $d = (n-1)p$. If %there exists $\delta > 0$ such that
\begin{equation}\label{conditionsforERtheoremSDP}
\frac{d}{\log n} \geq (1+\delta)\left( \frac4{(1-2\varepsilon)^2} + \frac4{3(1-2\varepsilon)}\right),
\end{equation}
then the SDP achieves exact recovery with probability at least $1-n^{-\delta}$.
When $\epsilon \to \frac12$, condition (\ref{conditionsforERtheoremSDP}) is equivalent to
\begin{equation}\label{conditionSDPepsilon12}
\frac{d}{\log n} \geq 4\frac{(1+\delta)}{(1-2\varepsilon)^2} + o\left(\frac1{(1-2\varepsilon)^2}\right).
\end{equation}
\end{theorem}
Note that, when $\epsilon \to \frac12$, condition (\ref{conditionSDPepsilon12}) differs from (\ref{eq:suffCondErdRenBigEps}), the sufficient condition for exact recovery with the maximum likelihood estimator, by a multiplicative factor of $2$. This gap is further discussed in Section~\ref{section:directions}.

%%%%%%%%%%%%%%%%%%%%%%%%%%%%%%%%%%%%%%%%%%%%%%%%%%
%%%%%%%%%%%%%%%%%%%%%%%%%%%%%%%%%%%%%%%%%%%%%%%%%%
%%%%%%%%%%%%%%%%%%%%%%%%%%%%%%%%%%%%%%%%%%%%%%%%%%
\subsection{Deterministic regular graph}

We now treat the case in which the underlying graph is a deterministic $d$-regular graph $G=(V,E)$ and use condition~(\ref{conditionneededdual}) to give guarantees for exact recovery.

%\[
%\lambda_2 = \frac1d \min_{x\perp 1}\left(x^TL_Gx\right) \text{ and } \lambda_n = \frac1d \max_x \left(x^TL_Gx\right) 
%\]
We need a measure of connectivity for $G$. Let $A_G = dI_{n\times n} - L_G$ be the adjacency matrix of $G$, let $\lambda_2$ be the second largest eigenvalue of $\frac1dA_G$, and let $\lambda_n$ be the smallest eigenvalue of $\frac1dA_G$. Since $G$ has no self-loop we have $\lambda_n<0$, which means
\begin{equation}\label{def:lambda2andlambdanforG}
\lambda_2 = \frac1d \max_{x\perp 1}\left( \frac{x^TA_Gx}{x^Tx} \right)\text{ and } |\lambda_n| = \frac1d \max_{x\perp 1} \left( - \frac{x^TA_Gx}{x^Tx} \right).
\end{equation}

This immediately gives $\lambda'_{\min}(L_G) =  d(1-\lambda_2)$ and $\lambda_{\max}(L_G)\leq d(1+|\lambda_n|)$, where $\lambda'_{\min}(\cdot)$ does not take into account the subspace generated by $1$.

As in the previous section, for each edge $e$ incident in the pair of vertices $i<j$, let $\Lambda_{e}$ be the matrix that is $1$ in the entries $(i,i)$ and $(j,j)$, $-1$ in the entries $(i,j)$ and $(j,i)$, and $0$ elsewhere. Observe that $\Lambda_{ij}\succeq 0$ and
\(
 L_G = \sum_{e\in E} \Lambda_e.
\)

Given $e\in E$, let $\alpha_{e}$ be the random variable that takes the value $1$ if edge $e$ is not in $H$ and the value $-1$ if it is in $H$. Hence $\alpha_{e}$ are i.i.d.\ and take the values $1,-1$ with probability
 \[
  \alpha_e = \left\{ \begin{array}{cl}  
     1 & \text{ with probability }  \left(1-\varepsilon\right) \\ 
     -1 & \text{ with probability }  \varepsilon.
       \end{array}   \right.
 \]

In the new notation,
\(
 L_G - 2L_H = \sum_{e\in E} \alpha_e\Lambda_{e}.
\)

Recall that we want to understand when there exists $\delta>0$ for which:
\begin{equation}\label{generalgraphdelta123}
 \Prob\left[ L_G - 2L_H\succeq 0 \right] \geq 1 - n^{-\delta}.
\end{equation}

As before, let us consider the centered variables $A_{e} = \left( 1-2\eps - \alpha_e \right)\Lambda_e$ and $A = \sum_{e\in E}A_e$. We have
\[
L_G - 2L_H = (1-2\eps)\sum_{e\in E}\Lambda_e - A = (1-2\eps)L_G - A.
\]

This means that $\lambda_{\max}(A) \leq (1-2\eps)\lambda_{\min}(L_G)$ is a sufficient condition for $L_G - 2L_H\succeq 0$. Since $\lambda_{\min}(L_G)\geq d(1-\lambda_2)$,
\[
\lambda_{\max}(A) \leq d(1-2\eps)(1-\lambda_2)
\]
is also sufficient.

Just as in the Section above, we use the Matrix Bernstein inequality (Theorem 1.4 in~\cite{Tropp:TailBoundsRM}), which implies
\(
 \Prob\left[ \lambda_{\max}(A) \geq t  \right] \leq n \exp\left( - \frac{t^2/2}{\sigma^2+Rt/3} \right),
\)
where $\sigma^2 = \left\| \sum_{e\in E} \EE A_{e}^2 \right\|$, with $\|\cdot\|$ denoting the spectral norm, and $R\geq\lambda_{\max}\left(A_e\right)$. This means that:
\begin{align*}
\sigma^2 & = \EE(1-2\eps - \alpha_e)^2\left\|   \sum_{e\in E}\Lambda_{e}^2 \right\| \\
&= (4\eps(1-\eps)) 2\lambda_{\max}(L_G) \leq 8\eps(1-\eps)d(1+|\lambda_n|),
\end{align*}
and we can take
\(
R = 4(1-\eps).
\)
Plugging everything together,
\begin{align*}
 &\Prob\left[ \lambda_{\max}(A) \geq t  \right] \\
 & \quad \leq n \exp\left( - \frac{t^2/2}{8\eps(1-\eps)d(1+|\lambda_n|)+4(1-\eps)t/3} \right).
\end{align*}

Setting $t = d(1-2\eps)(1-\lambda_2)$ gives,

$\Prob\left[ \lambda_{\max}(A) \geq d(1-2\eps)(1-\lambda_2)  \right]  $ 
$\quad~\quad~\leq~n~\exp\left( - d\frac{(1-2\eps)^2(1-\lambda_2)^2}{16\eps(1-\eps)(1+|\lambda_n|)+\frac83(1-\eps)(1-2\eps)(1-\lambda_2)} \right)$.

This means that it suffices to have

$n \exp\left( - d\frac{(1-2\eps)^2(1-\lambda_2)^2}{16\eps(1-\eps)(1+|\lambda_n|)+\frac83(1-\eps)(1-2\eps)(1-\lambda_2)} \right) \leq n^{-\delta},$
which is equivalent to
%\begin{eqnarray*}
% \frac1d(1+\delta)\log n & \leq &  \frac{(1-2\eps)^2(1-\lambda)^2}{16\eps(1-\eps)(1+\lambda)+\frac83(1-\eps)(1-2\eps)(1-\lambda)} \\
% d & \geq &  \frac{16\eps(1-\eps)(1+\lambda)+\frac83(1-\eps)(1-2\eps)(1-\lambda)}{(1-2\eps)^2(1-\lambda)^2}(1+\delta)\log n \\
% d & \geq &  \frac{16\eps(1-\eps)(1+\lambda)}{(1-2\eps)^2(1-\lambda)^2}+\frac{\frac83(1-\eps)(1-2\eps)(1-\lambda)}{(1-2\eps)^2(1-\lambda)^2}(1+\delta)\log n\\
% d & \geq &  \left[16\frac{\eps(1-\eps)}{(1-2\eps)^2}+\frac83\frac{(1-\eps)(1-\lambda)}{(1-2\eps)(1+\lambda)}\right]\frac{1+\lambda}{(1-\lambda)^2}(1+\delta)\log n.
%\end{eqnarray*}

%\begin{equation}\label{expressiondregular453}

$d  \geq   \left[16\frac{\eps(1-\eps)}{(1-2\eps)^2}+\frac83\frac{(1-\eps)(1-\lambda_2)}{(1-2\eps)(1+|\lambda_n|)}\right]\frac{1+|\lambda_n|}{(1-\lambda_2)^2}(1+\delta)\log n$
%\end{equation}

Since $\eps(1-\eps) = \frac14 - \frac{(1-2\eps)^2}4$ and $1-\eps = \frac12 + \frac12(1-2\eps)$, we can rewrite  the expression above %(\ref{expressiondregular453})
 as $d/(1+\delta) \geq $
 
$ \left[\frac{1}{(1-2\eps)^2}-1+\frac13\left(1+ \frac1{1-2\eps} \right)\frac{1-\lambda_2}{1+|\lambda_n|}\right]4\frac{1+|\lambda_n|}{(1-\lambda_2)^2}\log n$, 

which concludes the proof of the main result of this section.
\begin{theorem}~\label{SDPtheoremfordregular}
 Let $G$ be a $d$-regular graph, and let $\lambda_2$ and $\lambda_n$ be defined as in (\ref{def:lambda2andlambdanforG}). As long as 
\begin{align}
 &\frac{d}{\log n}  \geq  4\frac{1+|\lambda_n|}{(1-\lambda_2)^2}(1+\delta)   \times \\ 
 &\times \left[\frac{1}{(1-2\eps)^2}+\frac13\frac{1-\lambda_2}{(1-2\eps)(1+|\lambda_n|)}+\frac13\frac{1-\lambda_2}{(1+|\lambda_n|)}-1\right], &\nonumber
\end{align}
the SDP achieves exact recovery with probability at least $1-n^{\delta}$.

Moreover, if $\eps\to \frac12$, this can be rewritten as
\begin{equation*}
 \frac{d}{\log n}  \geq  4\frac{1+|\lambda_n|}{(1-\lambda_2)^2}(1+\delta) \left[\frac{1}{(1-2\eps)^2} +o\left(\frac1{(1-2\eps)^2}\right)\right].
\end{equation*}
If, furthermore, $\lambda_2 = o(1)$ and $|\lambda_n| = o(1)$ the condition reads:
\begin{equation}
 \frac{d}{\log n}  \geq  4(1+\delta)\frac{1}{(1-2\eps)^2} +o\left(\frac1{(1-2\eps)^2}\right).
\end{equation}
\end{theorem}

 \begin{remark}
The case where $\max\{\lambda_2,|\lambda_n|\} = o(1)$ is of particular interest as this is satisfied for random $d$-regular graphs as, for every $\delta>0$, $\max\{\lambda_2,|\lambda_n|\} \leq 2\frac{\sqrt{d-1}+\delta}d$ with high probability~\cite{Puder_ExpansionRandomGraphs,Friedman_randomdregularexpansion}. Also, if $G$ is a $d$-regular Ramanujan expander, then $\max\{\lambda_2,|\lambda_n|\} \leq 2\frac{\sqrt{d-1}}d$.   \end{remark}

Theorem~\ref{SDPtheoremfordregular} and Theorem~\ref{IT_sufficient_CheegerConstant} can be compared using Cheeger's inequality.
\begin{theorem}[Cheeger's inequality~\cite{NAlon_1986,NAlon_VMilman_1986}]\label{thm:Cheegersineq}
Let $G$ be a $d$-regular graph and let $h_G$ be its Cheeger constant (see~(\ref{def:Cheegerconstant44})) and $\lambda_2$ as defined in (\ref{def:lambda2andlambdanforG}), then
\begin{equation}\label{Cheeger_ineq}
\frac{1-\lambda_2}{2}\leq h_G \leq \sqrt{2\left(1-\lambda_2\right)}.
\end{equation}
\end{theorem}
Using (\ref{Cheeger_ineq}) it is easy to see that (when $\eps \to \frac12$) the IT sufficient condition (\ref{eq:cheegSuffCond_23}) in Theorem~\ref{IT_sufficient_CheegerConstant} is implied by
\[
\frac{d}{\log n} > \frac{4}{(1-\lambda_2) \left( 1 - 2 \varepsilon\right)^2} + o \! \left( \frac{1}{(1-\lambda_2) \left( 1 - 2 \varepsilon\right)^2} \right).
\]

\subsection{An alternative method based on $2$-length path voting}\label{Section:Andreasmethod}

In this Section we analyse a simple method to recover the vertex-variables based on $2$-length path voting. This method was proposed to the authors by Andrea Montanari, we thank Andrea for allowing us to analyse the method in this paper.

We will consider the Erd\H{o}s-R\'{e}nyi model. Let $G$ be drawn from the Erd\H{o}s-R\'{e}nyi distribution with parameters $n$ and $p$ and $\eps$ be the probability of an edge being incorrect. The recovery algorithm consists of: first one picks a center node, sets it to $1$, and then sets the value of every other node by looking at all paths of length $2$ between this node and the center node and by taking majority-voting among those.

In order to analyse the method, let us assume that the center node has been picked. For each of the other nodes, there are $n-2$ possible 2-length paths (corresponding to each one of the other $n-2$ vertices). For each of these vertices let us define the random variable $Y_k$ to be $0$ if there is no path, $-1$ if the path gives the wrong answer and $1$ if it gives the correct one. This means that the random variables $Y_k$ are i.i.d. and distributed as
\[
 Y_k = \left\{ \begin{array}{cl}  
   0 & \text{ with probability } 1-{p^2} \\
    -1 & \text{ with probability } {p^2} 2 \eps \left(1-\eps\right) = {p^2} \left[2\eps - 2\eps^2\right]\\ 
    %1 & \text{ with probability } {p^2} \left[\eps^2+\left(1-\eps\right)^2\right] = {p^2}\left[1 - 2\eps + 2\eps^2\right].
      1 & \text{ with probability }  {p^2}\left[1 - 2\eps + 2\eps^2\right].
      \end{array}   \right.
\]

The voting scheme succeeds for that one node as long as $\sum_{k=1}^{n-2}Y_k>0$.

Since we want to union-bound over $n-1$ vertices, and we want recovery to hold with probability at least $n^{-\delta}$, we want to understand for which $p$ and $\eps$ we have

\[
\Prob\left[ \sum_{k=1}^{n-2}Y_k \leq 0\right] \leq \frac1{n^{1+\delta}} \leq \frac1{(n-1)n^\delta}.
\]
Let us define the centered variable
% \begin{eqnarray*}
% X_k & = &  Y_k - \EE Y_k = Y_k + {p^2}\left( - 1 + 4\eps - 4\eps^2 \right) \\ & = & Y_k - {p^2} \left[1 - 4\eps(1-\eps)\right] = Y_k - {p^2} (1-2\eps)^2.
% \end{eqnarray*}
\[
 X_k  = Y_k - \EE Y_k = Y_k - {p^2} (1-2\eps)^2.
\]

This means we are interested in understanding when
\[
\Prob\left[ \sum_{k=1}^{n-2}X_k \leq -(n-2){p^2}(1-2\eps)^2\right] \leq \frac1{n^{1+\delta}},
\]
where $X_k = Y_k - {p^2}(1-2\eps)^2 $ is centered with distribution
\[
 X_k = \left\{ \begin{array}{cl}  
   -{p^2}(1-2\eps)^2 & \text{ with prob. } 1-{p^2} \\
    %-1-{p^2}(1-2\eps)^2 & \text{ with probability } {p^2}2\eps\left(1-\eps\right) = {p^2} \left[2\eps - 2\eps^2\right]\\ 
    %1-{p^2}(1-2\eps)^2 & \text{ with probability } {p^2} \left[\eps^2+\left(1-\eps\right)^2\right] = {p^2}\left[1 - 2\eps + 2\eps^2\right].
    -1-{p^2}(1-2\eps)^2 & \text{ with prob. }  {p^2} \left[2\eps - 2\eps^2\right]\\ 
    1-{p^2}(1-2\eps)^2 & \text{ with prob. } {p^2}\left[1 - 2\eps + 2\eps^2\right].
      \end{array}   \right.
\]
Also $|X_k|\leq 1+{p^2}(1-2\eps)^2$ and 
\begin{eqnarray*}
 \EE X_k^2 & = & (1-{p^2})\left({p^2}(1-2\eps)^2\right)^2  \\ & & + {p^2}\left[2\eps - 2\eps^2\right]\left(1+{p^2}(1-2\eps)^2\right)^2 \\ & & + {p^2}\left[1 - 2\eps + 2\eps^2\right]\left(1-{p^2}(1-2\eps)^2\right)^2 \\  & \leq & {p^2}.
\end{eqnarray*}
Bernstein's inequality thus gives
\begin{align*}
&\Prob\left[ \sum_{k=1}^{n-2}X_k  \leq  -t \right] \\  &\quad \leq   \exp\left(-\frac{t^2/2}{(n-2)\EE X_k^2 + \frac13\sup|X_k|t} \right) \\
 &\quad \leq  \exp\left(-\frac{t^2/2}{(n-2){p^2} + \frac13(1+{p^2}(1-2\eps)^2)t} \right).
\end{align*}
Replacing $t$ by $ (n-2){p^2}(1-2\eps)^2$ one gets
\begin{align*}
 &\Prob\left[ \sum_{k=1}^{n-2}X_k \leq -(n-2){p^2}(1-2\eps)^2\right] \\
& \quad \leq \exp\left(-\frac{(n-2){p^2}(1-2\eps)^4/2}{1 + \frac13(1+{p^2}(1-2\eps)^2)(1-2\eps)^2} \right).
\end{align*}
%\begin{eqnarray*}
% \Prob\left[ \sum_{k=1}^{n-2}X_k \leq -(n-2){p^2}(1-2\eps)^2\right] & \leq & \exp\left(-\frac{\left[(n-2){p^2}(1-2\eps)^2\right]^2/2}{(n-2){p^2} + \frac13(1+{p^2}(1-2\eps)^2)\left[(n-2){p^2}(1-2\eps)^2\right]} \right) \\
%& = & \exp\left(-\frac{(n-2){p^2}(1-2\eps)^4/2}{1 + \frac13(1+{p^2}(1-2\eps)^2)(1-2\eps)^2} \right).
%\end{eqnarray*}
This condition can be rewritten as,
\[
 \frac{(n-2){p^2}(1-2\eps)^4/2}{1 + \frac13(1+{p^2}(1-2\eps)^2)(1-2\eps)^2} \geq (1+\delta)\log n.
\]
In particular, when $\eps \to \frac12$, the sufficient condition can be written as
%\begin{equation}\label{conditionsforAndreasmethod3}
\[
 \frac{d^2}{\log n} \geq  2(1+\delta)\left( \frac1{(1-2\eps)^4} + o\left( \frac1{(1-2\eps)^4}\right)\right)n,
\]
%\end{equation}
where $d = pn$ is the expected average degree. Finally, we rewrite it in terms of $\frac{d}{\log n}$:
\begin{equation}\label{conditionsforAndreasmethod3}
 \frac{d}{\log n} \geq  \sqrt{2(1+\delta)}\left( \frac1{(1-2\eps)^2} + o\left( \frac1{(1-2\eps)^2}\right)\right)\sqrt{\frac{n}{\log n}}.
\end{equation}
Note that condition (\ref{conditionsforAndreasmethod3}) is asymptotically worse than the one obtained for the SDP-based approach (Theorem~\ref{SDP_ErdosReinyi}). In particular, it forces the average degree to be at least of order $\sqrt{n}$.

\section{Directions and open problems}\label{section:directions}

There are various extensions to consider for the above models, including the generalization to $q$-ary instead of binary variables and the extension to problems with hyperedges instead of edges as in \cite{random}. Non-binary variables would be particularly interesting for the synchronization problem in higher dimension, where the orthogonal matrices are quantized to a higher order. There are several extensions that are interesting for applications in community detection. First, it would be important to investigate non-symmetric noise models, i.e., noise models that are non-additive. First steps towards this were recently taken in~\cite{Abbe_ExactSBM}. Then, it would be interesting to study partial (as opposed to exact) recovery for sparse graphs with constant degrees, or to incorporate constraints on the size of the communities. In particular, it would be interesting to analyze the behavior of the SDP approach in the partial recovery regime, as it would potentially require a rounding step. One can also extend the family of base-graph ensembles. A particularly interesting future direction is to investigate characteristics of deterministic graphs that can provide IT lower-bounds for recovery. As we have seen, the lack of spectral gap alone is insufficient for that purpose.

\begin{figure}[h]
\label{Phasetransition_pic_1}
\begin{center}
\includegraphics[width=0.45\textwidth, height=0.24\textheight]
{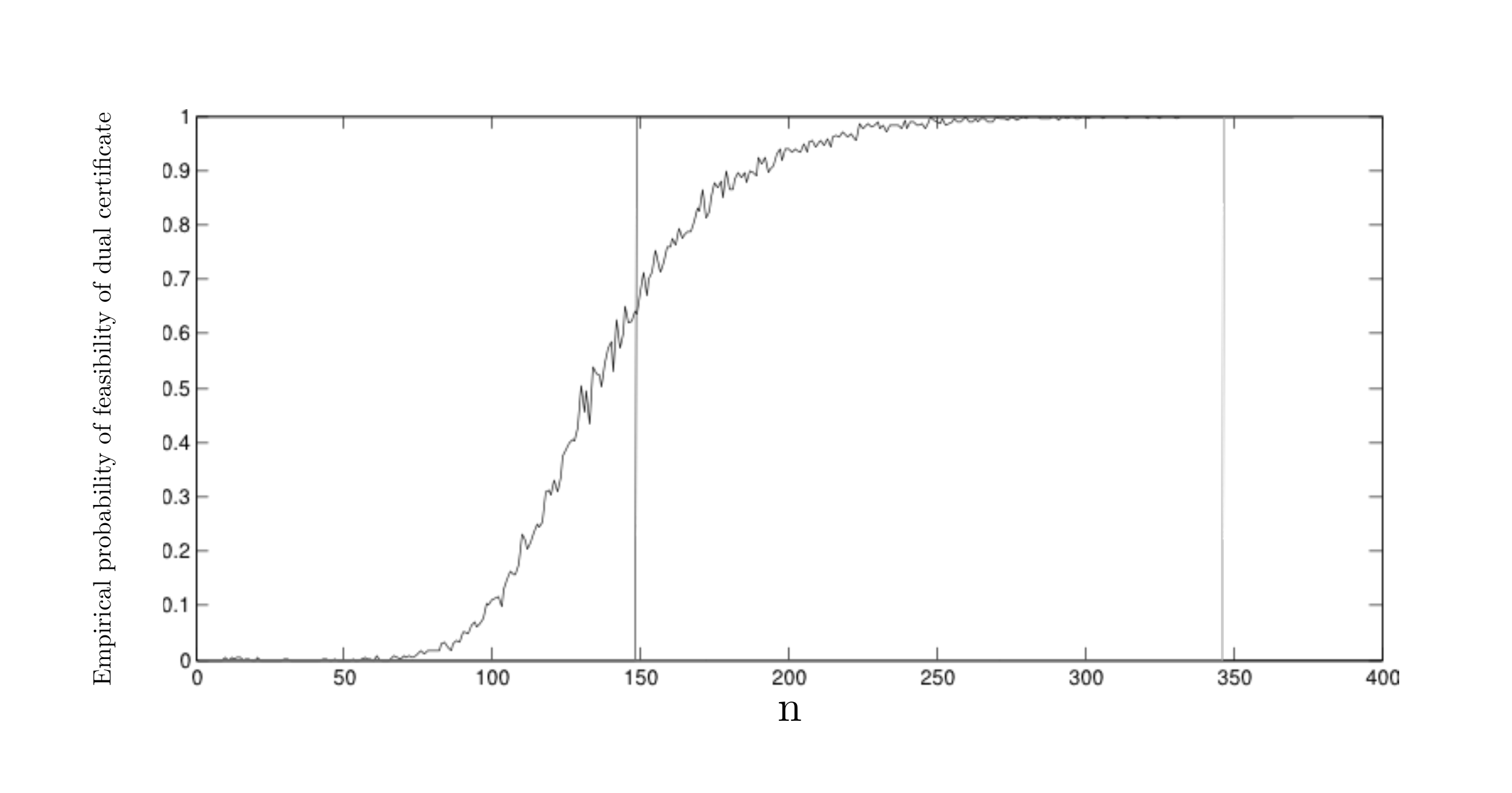}
\linebreak
\includegraphics[, width=0.45\textwidth, height=0.24\textheight]
{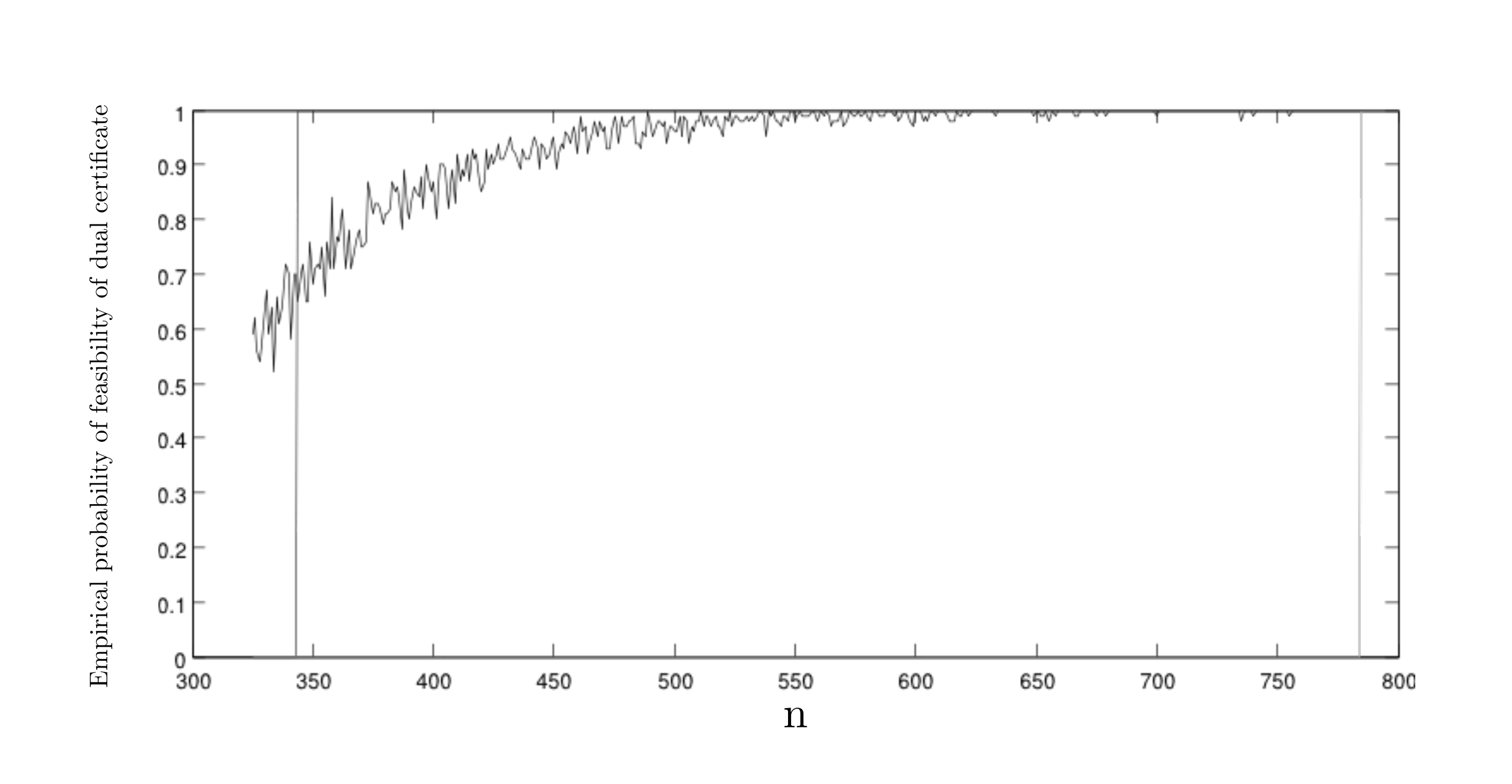}
\caption{{\small  Results of a simple simulation where, given the edge probability parameter $p$ and noise level parameter $\epsilon$, we generated random instances of the problem for different values of the number of vertices $n$ and checked whether the dual certificate proposed, $L_G-2L_H$, is PSD. The plot shows (on the y-axis), for different values of $n$ (on the x-axis), the ratio of trials that have a PSD dual certificate. The two vertical lines correspond to the thresholds of the IT and the SDP guarantees. The plot on the top is constructed with $p = 0.75$, $\eps = 0.35$, and the experiment is run $500$ times for each value of $n$. The plot on the bottom is constructed with $p = 0.85$, $\eps = 0.4$, and the experiment is run $100$ times for each value of $n$. }\normalsize}
\end{center}
\end{figure}

Finally, it would be interesting to better understand the gap between the IT rates and the ones we showed for our SDP-based algorithm. With this in mind, we ran a simple simulation where, given $p$ and $\epsilon$, we generated random instances of the problem for different values of $n$ and checked whether the dual certificate proposed was feasible. The results, reported in Figure~\ref{Phasetransition_pic_1}, suggest that this does not happen all the way down to the IT threshold suggesting that the gap might be a shortcoming of the method and not an artefact of the analysis. However, it is possible that a sharper analysis can yield better guarantees for the SDP-based algorithm. In particular, our analysis hinges on an all-purpose matrix Bernstein inequality that may be suboptimal in this case, and a specialized study of the particular random matrix in question may yield better results. We defer such a study for future investigations. Although the fact that the dual certificate is not feasible does not necessarily imply that the SDP is not achieving exact recovery, checking the dual certificate is considerably cheaper from a computational point of view, and other experiments, not reported, showed that the two tests are essentially equivalent in practice. This poses the natural question of whether there exists a polynomial-time algorithm that is able to match the rates achieved by the ML estimator. The existence of a gap between the performance of the ML estimator and the best polynomial-time algorithm would be extremely interesting.

%%%%%%%%%%%%%%%%%%%%%%%%%%%%%%%%%%%%%%%%%%%%%%%%%%%%%%%%%%%%%%%%%%%%%%%%%%

\section*{Acknowledgements}
We thank Andrea Montanari for suggesting to us the algorithm described in Section~\ref{Section:Andreasmethod} and Joel Tropp for insightful discussions regarding~\cite{Tropp:TailBoundsRM}.

We would also like to thank Yuxin Chen, Andrea Goldsmith, Peter Huang,
and Leo Guibas for useful discussions and for making their related, then
unpublished, work~\cite{Chen_Goldsmith_ISIT2014,Chen_Huang_Guibas_Graphics} available to us following the announcement of
the results of this paper by ASB at a seminar in Stanford University. It
would otherwise have been impossible for us to appropriately address their results
in this paper.

A. S. Bandeira was supported by AFOSR Grant No. FA9550-12-1-0317.
A. Singer was partially supported by Award Number R01GM090200 from the NIGMS, by Award Number FA9550-12-1-0317 and FA9550-13-1-0076 from AFOSR,
and by Award Number LTR DTD 06-05-2012 from the Simons Foundation.

\bibliographystyle{plain}
\bibliography{ExactRecovery.bib,gen.bib}

\end{document}